\documentclass[preprint,aps,showpacs,nofootinbib,preprintnumbers,amsmath,amssymb]{revtex4-1}
\usepackage{epsfig}
\usepackage{subfigure}
\usepackage{dcolumn}
\usepackage{bm}
\usepackage[usenames ,dvipsnames]{xcolor}
\usepackage{slashed}
\usepackage{graphicx,color}

\begin{document}
\title{Charmed Baryon Weak Decays with SU(3) Flavor Symmetry}

\author{C.Q. Geng$^{1,2}$, Y.K. Hsiao$^{1,2}$, Chia-Wei Liu$^{2}$ and Tien-Hsueh Tsai$^{2}$}
\affiliation{
$^{1}$School of Physics and Information Engineering, Shanxi Normal University, Linfen 041004, China\\
$^{2}$Department of Physics, National Tsing Hua University, Hsinchu, Taiwan 300
}\date{\today}

\begin{abstract}
We study the semileptonic and non-leptonic charmed baryon decays
with  $SU(3)$ flavor symmetry, where the charmed baryons can be
${\bf B}_{c}=(\Xi_c^0,\Xi_c^+,\Lambda_c^+)$,
${\bf B}'_{c}=(\Sigma_c^{(++,+,0)},\Xi_{c}^{\prime(+,0)},\Omega_c^0)$,
${\bf B}_{cc}=(\Xi_{cc}^{++},\Xi_{cc}^+,\Omega_{cc}^+)$, or
${\bf B}_{ccc}=\Omega^{++}_{ccc}$.
With ${\bf B}_n^{(\prime)}$ denoted as the baryon octet (decuplet),
we find that the ${\bf B}_{c}\to {\bf B}'_n\ell^+\nu_\ell$ decays
are forbidden, while the
$\Omega_c^0\to \Omega^-\ell^+\nu_\ell$,
$\Omega_{cc}^+\to\Omega_c^0\ell^+\nu_\ell$, and
$\Omega_{ccc}^{++}\to \Omega_{cc}^+\ell^+\nu_\ell$ decays are
the only existing Cabibbo-allowed modes
for ${\bf B}'_{c}\to {\bf B}'_n\ell^+\nu_\ell$, ${\bf B}_{cc}\to {\bf B}'_c\ell^+\nu_\ell$,
and ${\bf B}_{ccc}\to {\bf B}_{cc}^{(\prime)}\ell^+\nu_\ell$,
respectively.
We predict the rarely studied
${\bf B}_{c}\to {\bf B}_n^{(\prime)}M$ decays, such as
${\cal B}(\Xi_c^0\to\Lambda^0\bar K^0,\,\Xi_c^+\to\Xi^0\pi^+)
=(8.3\pm 0.9,8.0\pm 4.1)\times 10^{-3}$
and ${\cal B}(\Lambda_c^+\to \Delta^{++}\pi^-,\,\Xi_c^0\to\Omega^- K^+)
=(5.5\pm 1.3,4.8\pm 0.5)\times 10^{-3}$.
For the observation, the doubly and triply charmed baryon decays of
$\Omega_{cc}^{+}\to \Xi_c^+\bar K^0$,
$\Xi_{cc}^{++}\to (\Xi_c^+\pi^+$, $\Sigma_c^{++}\bar K^0)$,
and $\Omega_{ccc}^{++}\to (\Xi_{cc}^{++}\bar K^0,\Omega_{cc}^+\pi^+,\Xi_c^+ D^+)$
are the favored Cabibbo-allowed decays,
which are accessible to the BESIII and LHCb experiments.
\end{abstract}

\maketitle
\section{introduction}
Since 2016, the BESIII Collaboration has richly reanalyzed
the singly charmed baryon decays, such as
$\Lambda_{c}^+(2286) \to p \bar K^0, \Lambda \pi^+,\Sigma^+ \pi^0$ and
$\Sigma^0\pi^+$~\cite{pdg,Ablikim:2015flg},  with higher precision.
In addition, the Cabibbo-suppressed decays are measured
for the first time, where
${\cal B}(\Lambda_{c}^+(2286)\to p \eta)=(1.24\pm0.28\pm 0.10)\times10^{-3}$
and ${\cal B}(\Lambda_{c}^+(2286)\to p \pi^0)<3\times10^{-4}$ (90\% C.L.)~\cite{Ablikim:2017ors}.
On the other hand, the LHCb Collaboration has recently observed
the  decay of $\Xi_{cc}^{++}\to \Lambda_c K^-\pi^+\pi^+$~\cite{Aaij:2017ueg},
which is used to identify one of the doubly charmed baryon triplet,
$(\Xi_{cc}^{++},\Xi_{cc}^+,\Omega_{cc}^+)$, consisting of
$ccq$ with $q=(u,d,s)$, respectively.
These recent developments
suggest the possible measurements for the spectroscopy of
the singly, doubly and triply charmed baryons in the near future,
despite the not-yet-observed triply charmed baryon ones.
Moreover,
the charmed baryon formations and their decays would reveal
the underlying QCD effects,
which helps us to understand the recent discoveries of
the pentaquark and XYZ states that contain the charm quarks
also~\cite{Pc_LHCb,Pc_LHCb2,Aaij:2016ymb,QN_LHCb,Aaij:2014jqa}.

The spectroscopy of
the charmed baryons is built by measuring their decay modes.
For example,
the existence of the $\Xi_{cc}^+$ state was once reported
by the SELEX collaboration~\cite{Mattson:2002vu,Ocherashvili:2004hi},
but not confirmed by the other experiments~\cite{Ratti:2003ez,Aubert:2006qw,
Chistov:2006zj,Aaij:2013voa}.
%
Until very recently,
LHCb has eventually found the doubly charmed $\Xi_{cc}^{++}$ state at a mass of
$(3621.40 \pm 0.72\pm 0.27\pm 0.14)$~MeV~\cite{Aaij:2017ueg}, which is
reconstructed as the two-body $\Xi_{cc}^{++}\to \Sigma_c^{++}(2455)\bar K^{*0}$ decay
with the resonant strong decays of
$\Sigma_c^{++}\to \Lambda_c^+\pi^+$ and $\bar K^{*0}\to K^-\pi^+$, as shown by
 the theoretical calculation~\cite{Yu:2017zst}.
%
Note that the corresponding decay lifetime has not been determined yet.
It should be interesting to perform a full exploration of all possible charmed baryon decays,
and single out the suitable decay channels for the measurements.

To study the charmed baryon decays, since
the most often used factorization approach in the b-hadron decays~\cite{ali,Geng:2006jt,Hsiao:2014mua}
has been demonstrated not to work
for the two-body
${\bf B}_c\to {\bf B}_n M$ decays~\cite{Bjorken:1988ya,Lu:2016ogy},
where ${\bf B}_{n(c)}$ and $M$ are denoted as the (charmed) baryon and meson, respectively,
one has to compute the sub-leading-order contributions or the final state interactions
to take into account the non-factorizable effects~\cite{Cheng:1991sn,Cheng:1993gf,
Zenczykowski:1993hw,Fayyazuddin:1996iy,Dhir:2015tja}, 
whereas the QCD-based models in the ${\bf B}_c$ decays
are not available yet.
On the other hand, with the advantage of avoiding the
detailed dynamics of QCD, the approach with  $SU(3)$ flavor ($SU(3)_f$) symmetry
can relate decay modes in the $b$ and $c$-hadron decays~\cite{Lu:2016ogy,
He:2000ys,Fu:2003fy,Hsiao:2015iiu,He:2015fwa,He:2015fsa,
Savage:1989qr,Savage:1991wu,h_term,Wang:2017azm,Geng:2017esc},
where the $SU(3)$ amplitudes
receive  non-perturbative and non-factorizable effects,
despite the unknown sources.
In this paper,
in terms of  $SU(3)_f$ symmetry, we will examine
the semileptonic and non-leptonic two-body ${\bf B}_{c}$ decays,
search for  decay modes accessible to experiment,
and establish the spectroscopy of the charmed baryon states.
The analysis will explore
the consequences of neglecting a decay amplitude expected to be small.

Our paper is organized as follows.
In Sec. II, we develop the formalism,
where the Hamiltonians, (charmed) baryon and meson states
are presented in the irreducible forms under  $SU(3)_f$ symmetry.
The amplitudes of the semileptonic and non-leptonic decay modes
are given in Secs. III and IV, respectively.
In Sec. V, we discuss  all  possible decays and show the relationships among them as well as some numerical
results, which are relevant to the experiments.
We  conclude  in Sec. VI.

\section{Formalism}
\subsection{The effective Hamitonian}
For the semileptonic $c\to q \ell^+\nu_\ell$ transition with $q=(d,s)$,
the effective Hamiltonian at the quark-level is presented as
\begin{eqnarray}\label{Heff1}
{\cal H}_{eff}&=&\frac{G_F}{\sqrt 2}V_{cq}
(\bar qc)_{V-A} (\bar u_\nu v_\ell)_{V-A}\,,
\end{eqnarray}
where $G_F$ is the Fermi constant and
$V_{ij}$ are the Cabibbo-Kobayashi-Maskawa (CKM)
quark mixing matrix elements, while
$(\bar q_1 q_2)_{V-A}$ and $(\bar u_\nu v_\ell)_{V-A}$ stand for
$\bar q_1\gamma_\mu(1-\gamma_5)q_2$ and
$\bar u_\nu \gamma^\mu(1-\gamma_5)v_\ell$, respectively.
For the non-leptonic $c\to  s u\bar d$, $c\to u q\bar q$ and $c\to  du\bar s$ transitions,
one has the effective Hamiltonian to be
\begin{eqnarray}\label{Heff2}
\bar {\cal H}_{eff}&=&\frac{G_F}{\sqrt 2}
\{V_{cs}V_{ud}(c_+ O_+ +c_- O_-)
+V_{cd}V_{ud}(c_+\hat O_+ +c_- \hat O_-)
+V_{cd}V_{us}(c_+O'_+ +c_- O'_-)\},
\end{eqnarray}
with the four-quark operators $O_\pm^{(\prime)}$ and $\hat O_\pm\equiv O_\pm^d-O_\pm^s$ written as
\begin{eqnarray}\label{O12}
&&
O_\pm={1\over 2}[(\bar u d)_{V-A}(\bar s c)_{V-A}\pm (\bar s d)_{V-A}(\bar u c)_{V-A}]\,,\;\nonumber\\
&&
O_\pm^q={1\over 2}[(\bar u q)_{V-A}(\bar q c)_{V-A}\pm (\bar q q)_{V-A}(\bar u c)_{V-A}]\,,\;\nonumber\\
&&
O'_\pm={1\over 2}[(\bar u s)_{V-A}(\bar d c)_{V-A}\pm (\bar d s)_{V-A}(\bar u c)_{V-A}]\,,\;
\end{eqnarray}
where $V_{cd}V_{ud}=-V_{cs}V_{us}$ has been used.
According to $|V_{cd}V_{ud}|/|V_{cs}V_{ud}|=\sin\theta_c$ and
$|V_{cd}V_{us}|/|V_{cs}V_{ud}|=\sin^2\theta_c$ with $\theta_c$ known as the Cabibbo angle,
the operators for the $c\to  s u\bar d$, $c\to u q\bar q$ and $c\to  du\bar s$ transitions
represent the Cabibbo-allowed,
Cabibbo-suppressed and doubly Cabibbo-suppressed processes, respectively.
As the scale-dependent Wilson coefficients,
$c_\pm$ are calculated to be $(c_+,c_-)=(0.76,1.78)$ at the scale $\mu=1$ GeV
in the NDR scheme~\cite{Fajfer:2002gp,Li:2012cfa}.

Based on  $SU(3)_f$ symmetry,
the Lorentz-Dirac structures for the four-quark
 operators in Eq.~(\ref{O12}) are not explicitly expressed
with the quark index $q_i=(u,d,s)$ as an $SU(3)_f$ triplet $(3)$, such that  in Eq.~(\ref{Heff1})
the quark-current side of $(\bar qc)$ forms an anti-triplet ($\bar 3$), which leads to
\begin{eqnarray}\label{Heff1b}
{\cal H}_{eff}=\frac{G_F}{\sqrt 2}H(\bar 3)(\bar u_\nu v_\ell)_{V-A}\,,
\end{eqnarray}
with the tensor notation of $H(\bar 3)=(0,V_{cd},V_{cs})$, where
$V_{cs}=1$ and $V_{cd}=-\sin\theta_c$.
For the $c\to  s u\bar d$ and $c\to u q\bar q$ transitions in Eq.~(\ref{Heff2}),
the four-quark operators can be presented as
$(\bar q_i q^k)(\bar q_j c)$, with $\bar q_i q^k\bar q_j$ being decomposed as
$\bar 3\times 3\times \bar 3=\bar 3+\bar 3'+6+\overline{15}$.
Consequently,
the operators $O_{-,+}^{(\prime)}$ ($\hat O_{-,+})$ fall into the irreducible representations of
${\cal O}_{6,\overline{15}}^{(\prime)}$ ($\hat{\cal O}_{6,\overline{15}}$),
given by
\begin{eqnarray}
{\cal O}_6&=&{1\over 2}(\bar u d\bar s-\bar s d\bar u)c\,,\nonumber\\
{\cal O}_{\overline{15}}&=&{1\over 2}(\bar u d\bar s+\bar s d\bar u)c\,,\nonumber\\
\hat {\cal O}_6&=&{1\over 2}(\bar u d\bar d-\bar d d\bar u+\bar s s\bar u-\bar u s\bar s)c\,,\nonumber\\
\hat {\cal O}_{\overline{15}}&=&
{1\over 2}(\bar u d\bar d+\bar d d\bar u-\bar s s\bar u-\bar u s\bar s)c\,,\nonumber\\
{\cal O'}_6&=&{1\over 2}(\bar u s\bar d-\bar d s\bar u)c\,,\nonumber\\
{\cal O'}_{\overline{15}}&=&{1\over 2}(\bar u s\bar d+\bar d s\bar u)c\,,
\end{eqnarray}
which are in accordance with the tensor notations of
$H(6)_{ij}$ and $H(\overline{15})_i^{jk}$,
with the non-zero entries:
\begin{eqnarray}
&&H_{22}(6)=2\,,H_{23}(6)=H_{32}(6)=-2s_c \,,H_{33}(6)=2s_c^2\,,\nonumber\\
&&H_2^{13}(\overline{15})=H_2^{31}(\overline{15})=1\,,\nonumber\\
&&H_2^{12}(\overline{15})=H_2^{21}(\overline{15})=
-H_3^{13}(\overline{15})=-H_3^{31}(\overline{15})=s_c\,,\nonumber\\
&&{H}_3^{12}(\overline{15})={H}_{3}^{21}(\overline{15})=-s_c^2\,,
\end{eqnarray}
respectively, with $s_c\equiv \sin\theta_c$ to include
the CKM matrix elements into the tensor notations.
Accordingly, the effective Hamiltonian in Eq.~(\ref{Heff2}) is transformed as
\begin{eqnarray}\label{Heff2b}
\bar {\cal H}_{eff}&=&\frac{G_F}{\sqrt 2}[c_- H(6)+c_+ H(\overline{15})]\,,
\end{eqnarray}
where the contribution of $H(6)$ to
the decay branching ratio can be 5.5 times larger than that of
$H({\overline{15}})$ due to $(c_-/c_+)^2\simeq 5.5$.
The simplifications resulting from the neglect of the 15-plet will be investigated below.

\subsection{The (charmed) baryon states and mesons}
For the singly charmed baryon states, which consist of $q_1 q_2 c$ with
$q_1 q_2$ being decomposed as the irreducible representation of $3\times 3=\bar 3+6$,
there exist the charmed baryon anti-triplet and sextet, given by
\begin{eqnarray}\label{Bc1_state}
{\bf B}_{c}&=&(\Xi_c^0,\Xi_c^+,\Lambda_c^+)
\,,\nonumber\\
{\bf B}'_{c}&=&\left(\begin{array}{ccc}
\Sigma^{++}_{c}& \frac{1}{\sqrt{2}}\Sigma^{+}_{c} & \frac{1}{\sqrt{2}}\Xi'^{+}_{c}\\
 \frac{1}{\sqrt{2}}\Sigma^{+}_{c} &\Sigma^0_c  & \frac{1}{\sqrt{2}}\Xi'^{0}_{c}\\
 \frac{1}{\sqrt{2}}\Xi'^{+}_{c} & \frac{1}{\sqrt{2}}\Xi'^{0}_{c} &\Omega^0_c
\end{array}\right)\,,
\end{eqnarray}
respectively. Similarly, ${\bf B}_{cc}$ and ${\bf B}_{ccc}$ to consist of qcc and ccc represent
 the doubly charmed baryon triplet and triply charmed baryon singlet,
given by
\begin{eqnarray}
{\bf B}_{cc}&=&(\Xi_{cc}^{++},\Xi_{cc}^+,\Omega_{cc}^+)\,,\nonumber\\
{\bf B}_{ccc}&=&\Omega^{++}_{ccc}\,,
\end{eqnarray}
respectively.
The final states, ${\bf B}_n$, $M$ and $M_c$, being
 the lowest-lying baryon octet,  meson octet,
and the charmed meson anti-triplet, are written as
\begin{eqnarray}\label{B_8}
{\bf B}_n&=&\left(\begin{array}{ccc}
\frac{1}{\sqrt{6}}\Lambda+\frac{1}{\sqrt{2}}\Sigma^0 & \Sigma^+ & p\\
 \Sigma^- &\frac{1}{\sqrt{6}}\Lambda -\frac{1}{\sqrt{2}}\Sigma^0  & n\\
 \Xi^- & \Xi^0 &-\sqrt{\frac{2}{3}}\Lambda
\end{array}\right)\,,\nonumber\\
M&=&\left(\begin{array}{ccc}
\frac{1}{\sqrt{2}}\pi^0+\frac{1}{\sqrt{6}}\eta & \pi^- & K^-\\
 \pi^+ &-\frac{1}{\sqrt{2}}\pi^0+\frac{1}{\sqrt{6}}\eta& \bar K^0\\
 K^+ & K^0& -\sqrt{\frac{2}{3}}\eta
\end{array}\right)\,,\nonumber\\
M_c&=&(D^0,D^+,D^+_s)\,,
\end{eqnarray}
respectively.
We note that in our calculations,   $\eta$  is only considered as a member of an octet, without
treating it as an octet-singlet mixture to simplify the analysis.
%
In addition, we have the baryon decuplet, given by
\begin{eqnarray}\label{B_10}
&&{\bf B}'_n=\frac{1}{\sqrt{3}}
\left(\begin{array}{ccc}
\left(\begin{array}{ccc}
\sqrt{3}\Delta^{++}&\Delta^+ & \Sigma^{\prime +}\\
\Delta^+ &\Delta^0 & \frac{\Sigma^{\prime 0}}{\sqrt{2}}\\
\Sigma^{\prime +}& \frac{\Sigma^{\prime 0}}{\sqrt{2}}& \Xi^{\prime 0}
\end{array}\right),
\left(\begin{array}{ccc}
\Delta^+ &\Delta^0 & \frac{\Sigma^{\prime 0}}{\sqrt{2}}\\
\Delta^0 &\sqrt{3}\Delta^-& \Sigma^{\prime -}\\
\frac{\Sigma^{\prime 0}}{\sqrt{2}}&\Sigma^{\prime -}&\Xi^{\prime -}
\end{array}\right),
\left(\begin{array}{ccc}
\Sigma^{\prime +}& \frac{\Sigma^{\prime 0}}{\sqrt{2}}&\Xi^{\prime 0}\\
\frac{\Sigma^{\prime 0}}{\sqrt{2}}&\Sigma^{\prime -}&\Xi^{\prime -}\\
\Xi^{\prime 0}&\Xi^{\prime -}&\sqrt{3}\Omega^-
\end{array}\right)
\end{array}\right).
\end{eqnarray}

\section{Semileptonic charmed baryon decays}
In this section, we present the amplitudes for
the semileptonic
${\bf B}_{c}^{(\prime)}\to {\bf B}_n^{(\prime)}\ell^+\nu_\ell$,
${\bf B}_{cc}\to {\bf B}_{c}^{(\prime)}\ell^+\nu_\ell$, and
${\bf B}_{ccc}\to {\bf B}_{cc}\ell^+\nu_\ell$ decays under  $SU(3)_f$ symmetry.
In terms of ${\cal H}_{eff}$ in Eq.~(\ref{Heff1b}), the amplitudes of
${\cal A}({\bf B}_{c}^{(\prime)}\to {\bf B}_n^{(\prime)}\ell^+\nu_\ell)=
\langle {\bf B}_n^{(\prime)} \ell^+\nu_\ell|H_{eff}|{\bf B}_{c}^{(\prime)}\rangle$ are derived as
${\cal A}({\bf B}_{c}^{(\prime)}\to {\bf B}_n^{(\prime)}\ell^+\nu_\ell)=
\frac{G_F}{\sqrt 2}V_{cq}T({\bf B}_{c}^{(\prime)}\to {\bf B}_n^{(\prime)})(\bar u_\nu v_\ell)_{V-A}$,
where $T({\bf B}_{c}^{(\prime)}\to {\bf B}_n^{(\prime)})$ are given by
\begin{eqnarray}\label{semi_1}
&&T({\bf B}_{c}\to {\bf B}_n)=
\alpha_1 ({\bf B}_n)^i_jH^j(\bar 3)({\bf B}_{c})_i\,,
\nonumber\\
&&T({\bf B}'_{c}\to {\bf B}_n)=\alpha_2 ({\bf B}_n)^i_jH^l(\bar 3)({\bf B}'_{c})^{jk}\epsilon_{ilk}\,,
\nonumber\\
&&T({\bf B}'_{c}\to {\bf B}'_n)=\alpha_3 ({\bf B}'_n)_{ijk}H^i(\bar 3)({\bf B}'_{c})^{jk}\,,
%
\end{eqnarray}
with  $SU(3)$ parameters $\alpha_i$ ($i=1,2, 3$) associated with
the ${\bf B}_{c}^{(\prime)}\to {\bf B}_n^{(\prime)}\ell^+\nu_\ell$ decays.
Note that $T({\bf B}_{c}\to {\bf B}'_n)$ disappears in Eq.~(\ref{semi_1}).
This is due to the fact that
the symmetric baryon decuplet $({\bf B}'_n)_{ijk}$
and the anti-symmetric $\epsilon_{ijk}$ coexist
in the forms of
$({\bf B}'_n)_{ijk} H^i(\bar 3)({\bf B}_{c})_l\epsilon^{ljk}$ and
$({\bf B}'_n)_{ljk} H^i(\bar 3)({\bf B}_{c})_l\epsilon^{ijk}$,
which identically vanish~\cite{Savage:1989qr}.
We also obtain the $T$ amplitudes of
the ${\bf B}_{cc}\to {\bf B}_{c}^{(\prime)}\ell^+\nu_\ell$
and ${\bf B}_{ccc}\to {\bf B}_{cc}\ell^+\nu_\ell$ decays, given by
\begin{eqnarray}\label{semi_2}
&&
T({\bf B}_{cc}\to {\bf B}_{c})=
\beta_1  H_q^j(\bar 3)({\bf B}_{c})^k\epsilon_{ijk}({\bf B}_{cc})^i\,,
\nonumber\\
&&T({\bf B}_{cc}\to {\bf B}'_{c})=
\beta_2  H_q^j(\bar 3)({\bf B}'_{c})_{ij}({\bf B}_{cc})^i\,,
\nonumber\\
&&T({\bf B}_{ccc}\to {\bf B}_{cc})=\delta_1 ({\bf B}_{cc})_i H_q^i(\bar 3)\,,
\end{eqnarray}
with  $SU(3)$ parameters $\beta_{1,2}$ and $\delta_1$, where the subscript $q$ refers to the $d$ or $s$ quark
in ${\bf B}_{cc}$.
It is interesting to note that, for $T({\bf B}_{ccc}\to {\bf B}_{cc})$,
${\bf B}_{ccc}=\Omega_{ccc}^{++}$ as the charmed baryon singlet
has no $SU(3)$ flavor index to connect to the final states and Hamiltonian.
The full expanded $T$ amplitudes in Eqs.~(\ref{semi_1}) and (\ref{semi_2}),
corresponding to the semileptonic charmed baryon decays,
can be found in Table~\ref{tab_semi}.

\section{Non-leptonic charmed baryon decays}
To proceed, we start with the non-leptonic charmed baryon decays,
in which the charmed baryons are
the singly, doubly, and triply charmed baryon states,
${\bf B}_{c_i}=({\bf B}_{c}^{(\prime)},{\bf B}_{cc},{\bf B}_{ccc})$,
respectively.

\subsection
{The two-body ${\bf B}_{c}^{(\prime)}\to {\bf B}_n^{(\prime)}M$ decays}
In terms of  $SU(3)_f$ symmetry,
the amplitudes of the singly charmed
${\bf B}_{c}^{(\prime)}\to {\bf B}_n^{(\prime)}M$ decays
in the irreducible forms are derived as
\begin{eqnarray}\label{amp_SU3}
&&{\cal A}({\bf B}_{c}^{(\prime)}\to {\bf B}_n^{(\prime)} M)
=\langle {\bf B}_n^{(\prime)} M|\bar H_{eff}|{\bf B}_{c}^{(\prime)}\rangle
=\frac{G_F}{\sqrt 2}T({\bf B}_{c}^{(\prime)}\to {\bf B}_n^{(\prime)} M)\,,
\end{eqnarray}
where
\begin{eqnarray}\label{T_Bc1_1}
&&
T({\bf B}_{c}\to {\bf B}_n M)=\nonumber\\
&&
a_1 H_{ij}(6)T^{ik}({\bf B}_n)_k^l (M)_l^j+
a_2 H_{ij}(6)T^{ik}(M)_k^l ({\bf B}_n)_l^j+
a_3 H_{ij}(6)({\bf B}_n)_k^i (M)_l^j T^{kl}\nonumber\\
&&
+a_4({\bf B}_n)^k_l (M)^j_i H(\overline{15})^{li}_k ({\bf B}_{c})_j
+a_5({\bf B}_n)^i_j (M)^l_i H(\overline{15})^{jk}_l ({\bf B}_{c})_k\nonumber\\
&&
+a_6({\bf B}_n)^k_l (M)^i_j H(\overline{15})^{jl}_i ({\bf B}_{c})_k
+a_7({\bf B}_n)^l_i (M)^i_j H(\overline{15})^{jk}_l ({\bf B}_{c})_k\,,
\end{eqnarray}
\begin{eqnarray}\label{T_Bc1_1b}
&&
T({\bf B}_{c}\to {\bf B}_n^{\prime} M)=\nonumber\\
&&
a_8 ({\bf B}'_n)_{ijk}({\bf B}_{c})_l H_{nm}(6) (M)^i_o \epsilon^{jln}\epsilon^{kmo}+
a_9 ({\bf B}'_n)_{ijk} (M)^i_l H(\overline{15})_m^{jn}({\bf B}_{c})_n\epsilon^{klm}\nonumber\\
&&
+a_{10} ({\bf B}'_n)_{ijk} (M)^i_l H(\overline{15})_m^{jk}({\bf B}_{c})_n\epsilon^{lmn}+
a_{11} ({\bf B}'_n)_{ijk} (M)^l_m H(\overline{15})_l^{ij}({\bf B}_{c})_n\epsilon^{kmn}\,,
\end{eqnarray}
\begin{eqnarray}\label{T_Bc1_2}
&&
T({\bf B}'_{c}\to {\bf B}_n M)=\nonumber\\
&&
a_{12}H_{ij}(6)({\bf B}'_{c})^{ij}({\bf B}_n)^{l}_{k}(M)^{k}_{l}+
a_{13}H_{ij}(6)({\bf B}'_{c})^{kl}({\bf B}_n)^{i}_{k}(M)^{j}_{l}\nonumber\\
&&
+a_{14}H_{ij}(6)({\bf B}'_{c})^{jk}({\bf B}_n)^{l}_{k}(M)^{i}_{l}
+a_{15}H_{ij}(6)({\bf B}'_{c})^{jk}({\bf B}_n)^{i}_{l}(M)^{l}_{k}\nonumber\\
&&
+a_{16}({\bf B}_n)^{i}_{j}(M)^{k}_{l}H(\overline{15})_i^{jm}({\bf B}'_{c})^{ln}\epsilon_{kmn}
+a_{17}({\bf B}_n)^{i}_{j}(M)^{k}_{l}H(\overline{15})_i^{lm}({\bf B}'_{c})^{jn}\epsilon_{kmn}
\nonumber\\
&&
+a_{18}({\bf B}_n)^{m}_{n}(M)^{n}_{j}H(\overline{15})_k^{ij}({\bf B}'_{c})^{kl}\epsilon_{ilm}
+a_{19}({\bf B}_n)^{j}_{l}(M)^{k}_{n}H(\overline{15})_m^{il}({\bf B}'_{c})^{mn}\epsilon_{ijk}
\nonumber\\
&&
+a_{20}({\bf B}_n)^{j}_{n}(M)^{k}_{l}H(\overline{15})_m^{il}({\bf B}'_{c})^{mn}\epsilon_{ijk}\,,
\end{eqnarray}
and
\begin{eqnarray}\label{T_Bc1_2b}
&&
T({\bf B}'_{c}\to {\bf B}'_n M)=\nonumber\\
&&
a_{21}({\bf B}'_n)_{lkm}(M)^{i}_{n}H_{ij}(6)({\bf B}'_{c})^{lk}\epsilon^{jmn}+
a_{22}({\bf B}'_n)_{klm}(M)^{l}_{n}H_{ij}(6)({\bf B}'_{c})^{jk}\epsilon^{imn}\nonumber\\
&&
+a_{23}({\bf B}'_n)_{ijk}(M)^{m}_{l}H(\overline{15})_m^{lk}({\bf B}'_{c})^{ij}+
a_{24}({\bf B}'_n)_{ijk}(M)^{k}_{m}H(\overline{15})_l^{ij}({\bf B}'_{c})^{lm}\nonumber\\
&&
+a_{25}({\bf B}'_n)_{ijk}(M)^{l}_{m}H(\overline{15})_l^{ij}({\bf B}'_{c})^{km}
+a_{26}({\bf B}'_n)_{ijk}(M)^{j}_{l}H(\overline{15})_m^{kl}({\bf B}'_{c})^{im}\,,
\end{eqnarray}
with $T^{ij}\equiv\epsilon^{ijk}({\bf B}_{c})_k$. Note that the Wilson coefficients $c_\pm$
have been absorbed in  $SU(3)$ parameters  $a_i$, which can relate
all possible decay modes.
The full expansions of the $T$ amplitudes in Eqs.~(\ref{T_Bc1_1})-(\ref{T_Bc1_2b})
are given in Tables~\ref{tab_Bc1a}-\ref{tab_Bc1_f}.

\subsection
{The doubly charmed ${\bf B}_{cc}\to {\bf B}_n^{(\prime)}M_c$
and ${\bf B}_{cc}\to {\bf B}_{c}^{(\prime)}M$ decays}
In the doubly charmed baryon decays, the $T$ amplitudes of
${\bf B}_{cc}\to {\bf B}_n M_c$ and ${\bf B}_{cc}\to {\bf B}'_n M_c$ are written as
\begin{eqnarray}\label{T_Bc2_D}
&&
T({\bf B}_{cc}\to {\bf B}_n M_c)=\nonumber\\
&&
b_{1}({\bf B}_{cc})^{i}(M_c)^{j} ({\bf B}_n)^{k}_{j} H_{ik}(6)+
b_{2}({\bf B}_{cc})^{i}(M_c)^{j} ({\bf B}_n)^{k}_{i} H_{jk}(6)\nonumber\\
&&
+b_{3}({\bf B}_{cc})^{l}(M_c)^{i}({\bf B}_n)^{k}_{m}H(\overline{15})_l^{jm}\epsilon_{ijk}
+b_{4}({\bf B}_{cc})^{i}(M_c)^{l}({\bf B}_n)^{k}_{m}H(\overline{15})_l^{jm}\epsilon_{ijk}\,,
\end{eqnarray}
and
\begin{eqnarray}\label{T_Bc2_D2}
&&
T({\bf B}_{cc}\to {\bf B}'_n M_c)=\nonumber\\
&&
b_{5}({\bf B}_{cc})^{i}(M_c)^{j}({\bf B}'_n)_{iml}H(\overline{15})_j^{ml}+
b_{6}({\bf B}_{cc})^{i}(M_c)^{j}({\bf B}'_n)_{jml}H(\overline{15})_i^{ml}\,,
\end{eqnarray}
where ${\bf B}_n$ and ${\bf B}'_n$
represent the octet and decuplet of the the baryon states
in Eqs.~(\ref{B_8}) and (\ref{B_10}), respectively.
It is interesting to note that measuring
 the processes in Eq.~(\ref{T_Bc2_D2}) can be a test
of the smallness of the 15-plet.
For the ${\bf B}_{cc}\to  {\bf B}_{c}^{(\prime)} M$ decays,
the $T$ amplitudes are expanded as
\begin{eqnarray}\label{T_Bc2_Bc1}
&&
T({\bf B}_{cc}\to {\bf B}_{c} M)=\nonumber\\
&&
b_{7}({\bf B}_{cc})^{i}({\bf B}_{c})^{j}(M)^{k}_{i} H_{jk}(6)+
b_{8}({\bf B}_{cc})^{i}({\bf B}_{c})^{k}(M)^{j}_{k} H_{ij}(6)\nonumber\\
&&
+b_{9}({\bf B}_{cc})^{i}H(\overline{15})_l^{jk}({\bf B}_{c})^m(M)^{l}_{j}\epsilon_{ikm}
+b_{10}({\bf B}_{cc})^{l}H(\overline{15})_l^{jk}({\bf B}_{c})^i(M)^{m}_{j}\epsilon_{ikm}
\,,
\end{eqnarray}
and
\begin{eqnarray}\label{T_Bc2_Bc1b}
&&
T({\bf B}_{cc}\to {\bf B}'_{c} M)=\nonumber\\
&&
b_{11}({\bf B}_{cc})^{i}({\bf B}'_{c})_{jk}(M)^{l}_{i}H(\overline{15})_l^{jk}
+b_{12}({\bf B}_{cc})^{i}({\bf B}'_{c})_{jl}(M)^{k}_{i}H(\overline{15})_k^{jl}\nonumber\\
&&
+b_{13}({\bf B}_{cc})^{i}({\bf B}'_{c})_{jk}(M)^{k}_{l}H(\overline{15})_i^{jl}
+b_{14}({\bf B}_{cc})^{i}({\bf B}'_{c})_{ij}(M)^{k}_{l}H_{km}(6)\epsilon^{mjl}
\nonumber\\&&
+b_{15}({\bf B}_{cc})^{i}({\bf B}'_{c})_{jk}(M)^{k}_{l}H_{im}(6)\epsilon^{mjl}\,.
\end{eqnarray}
The full expansions of the $T$ amplitudes in Eqs.~(\ref{T_Bc2_D})-(\ref{T_Bc2_Bc1b})
are given in Tables~\ref{tab_B2c_a} and \ref{tab_B2c_b}.

\subsection
{The triply charmed ${\bf B}_{ccc}\to {\bf B}_{cc}M$
and ${\bf B}_{ccc}\to {\bf B}_{c}^{(\prime)}M_c$ decays}
For the triply charmed baryon decays,
there are three types of decay modes, that is,
${\bf B}_{ccc}\to {\bf B}_{cc}M$, ${\bf B}_{ccc}\to {\bf B}'_{c}M_c$,
and ${\bf B}_{ccc}\to {\bf B}_{c}M_c$.
The corresponding $T$ amplitudes are given by
\begin{eqnarray}\label{T_Bc3}
T({\bf B}_{ccc}\to {\bf B}_{cc} M)&=&
d_{1}({\bf B}_{cc})_i(M)^{j}_{k}H(\overline{15})_j^{ik}+
d_{2}({\bf B}_{cc})_i(M)^{j}_{k}H_{jl}(6)\epsilon^{ikl}\,,\nonumber\\
T({\bf B}_{ccc}\to {\bf B}'_{c} M_c)&=&
d_{3}({\bf B}'_{c})_{ij}(M_c)^{k}H(\overline{15})_k^{ij}\,,\nonumber\\
T({\bf B}_{ccc}\to {\bf B}_{c} M_c)&=&
d_{4}({\bf B}_{c})^i (M_c)^{j}H(6)_{ij}\,,
\end{eqnarray}
where ${\bf B}_{ccc}=\Omega_{ccc}^{++}$ as the charmed baryon singlet
has no $SU(3)$ flavor index to connect to
the final states and $H(6,\overline{15})$.
The full expansions of the $T$ amplitudes in Eq.~(\ref{T_Bc3})
are given in Table~\ref{tab_Bc3}.

\section{Discussions}

\subsection{Semileptonic charmed baryon decays}
By taking ${\cal B}(\Lambda_c^+\to\Lambda^0 e^+\nu_e)=
(3.6\pm 0.4)\times 10^{-2}$~\cite{pdg} as the experimental input,
and relating the possible ${\bf B}_{c}\to {\bf B}_n\ell^+\nu_\ell$ decays
with the $SU(3)$ parameter $\alpha_1$ in Table~\ref{tab_semi},
the branching ratios of the Cabibbo-allowed decays are predicted to be
\begin{eqnarray}\label{semi_pred}
{\cal B}(\Xi_{c}^{0}\to\Xi^{-}e^+\nu_e)&=&(11.9\pm 1.6)\times 10^{-2}\,,\nonumber\\
{\cal B}(\Xi_{c}^{+}\to\Xi^{0}e^+\nu_e)&=&(3.0\pm 0.5)\times 10^{-2}\,,
\end{eqnarray}
while the Cabibbo-suppressed ones are evaluated as
\begin{eqnarray}\label{semi_pred2}
{\cal B}(\Xi_{c}^{0}  \to  \Sigma^{-}e^+\nu_e)&=&(6.0\pm 0.8)\times 10^{-3}\,,\nonumber\\
{\cal B}(\Lambda_{c}^{+}  \to  ne^+\nu_e)&=&(2.7\pm 0.3)\times 10^{-3}\,,\nonumber\\
{\cal B}(\Xi_{c}^{+}  \to  \Sigma^{0}e^+\nu_e)&=&(0.8\pm 0.1)\times 10^{-3}\,,\nonumber\\
{\cal B}(\Xi_{c}^{+}  \to  \Lambda^{0}e^+\nu_e)&=&(2.5\pm 0.4)\times 10^{-4}\,,
\end{eqnarray}
where we have taken $(\tau_{\Xi_c^0},\tau_{\Xi_c^+},\tau_{\Lambda_c^+})=
(1.12^{+0.13}_{-0.10},4.42\pm 0.26,2.00\pm 0.06)\times 10^{-13}$ s
and $s_c=0.2248$~\cite{pdg}. Our result of
${\cal B}(\Lambda_{c}^{+}  \to  ne^+\nu_e)$ in Eq.~(\ref{semi_pred2}) agrees with
that in Ref.~\cite{Lu:2016ogy} by  $SU(3)_f$ symmetry also.
%
The ${\bf B}_{c}\to {\bf B}'_n \ell^+\nu_\ell$ decays are forbidden modes,
reflecting the fact that
the ${\bf B}_{c}$ and ${\bf B}'_n$ states are
the uncorrelated anti-symmetric triplet and symmetric decuplet, respectively,
which can be viewed as the interesting measurements
to test the broken symmetry.

In Table~\ref{tab_semi},
we illustrate the possible ${\bf B}'_{c}\to {\bf B}_n^{(\prime)}\ell^+\nu_\ell$ decays,
where ${\bf B}'_{c}$ stands for the singly charmed baryon sextet in Eq.~(\ref{Bc1_state}).
We remark that currently it is hard to observe
the weak decays with
${\bf B}'_{c}=(\Sigma_c^{++},\Sigma_c^{+},\Sigma_c^{0})$ and
${\bf B}'_{c}=(\Xi_{c}^{\prime+},\Xi_{c}^{\prime0})$,
as the $\Sigma_c$ and $\Xi'_c$ decays are
dominantly through the strong and electromagnetic interactions,
with  ${\cal B}(\Sigma_c\to \Lambda_c\pi)\approx 100\%$ and
$\Xi'_c\to \Xi_c\gamma$, respectively.
In contrast, the $\Omega_c^0$ state that decays weakly
can be measurable. In particular,
the $\Omega_c^0\to \Omega^-\ell^+\nu_\ell$ decay
with $\Omega^-=sss$ becomes the only possible
Cabibbo-allowed $\Omega_c^0$ case~\cite{Savage:1989qr},
whereas the $\Omega_c^0\to {\bf B}_n\ell^+\nu_\ell$ decays
with the baryon octet are forbidden. This is due to the fact that,
via the Cabibbo-allowed $c\to s\ell^+\nu_\ell$ transition,
the $\Omega_c^0$ baryon consists of $ssc$ transforms as the $sss$ state, and has
has no association with the the baryon octet.
In the Cabibbo-suppressed $css\to dss$ transition, one has
the $\Omega_c^0\to \Xi^{(\prime)-}\ell^+\nu_\ell$ decays
with $\Xi^{-}$ and $\Xi^{\prime-}$ from both baryon octet and decuplet.

For ${\bf B}_{cc}\to {\bf B}_{c}^{(\prime)}\ell^+\nu_\ell$,
it is found from Table~\ref{tab_semi} that
\begin{eqnarray}
\Gamma(\Xi_{cc}^{+}  \to  \Xi_{c}^{(\prime)0}\ell^+\nu_\ell)=
\Gamma(\Xi_{cc}^{++}  \to  \Xi_{c}^{(\prime)+}\ell^+\nu_\ell)\,,
\end{eqnarray}
which respect the isospin symmetry.
%
Like the singly charmed $\Omega_c^0$ cases,
the Cabibbo-allowed $\Omega_{cc}^+(ccs)\to css$ transition forbids
the $\Omega_{cc}^+\to {\bf B}_{c}\ell^+\nu_\ell$ decays,
but allows  $\Omega_{cc}^+\to\Omega_c^0\ell^+\nu_\ell$
with $\Omega_c^0=css$.
The Cabibbo-suppressed $\Omega_{cc}^+(ccs)\to cds$
transition permits  $\Omega_{cc}^+\to (\Xi^-_c,\Xi^{\prime-}_c)\ell^+\nu_\ell$.

In the ${\bf B}_{ccc}\to {\bf B}_{cc}\ell^+\nu_\ell$ decays,
 $SU(3)_f$ symmetry leads to two possible decay modes,
of which the branching ratios are related as
\begin{eqnarray}
s_c^2{\cal B}(\Omega_{ccc}^{++}\to \Omega_{cc}^+\ell^+\nu_\ell)=
{\cal B}(\Omega_{ccc}^{++}\to \Xi_{cc}^+\ell^+\nu_\ell)\,,
\end{eqnarray}
suggesting that the Cabibbo-allowed
$\Omega_{ccc}^{++}\to \Omega_{cc}^+\ell^+\nu_\ell$ decay
is more accessible to experiment.

\subsection{Non-leptonic charmed baryon decays}
\noindent
$\bullet$ The ${\bf B}_{c}\to {\bf B}_n^{(\prime)}M$ decays\\
In the $\Lambda_c^+\to {\bf B}_n M$ decays,
the PDG~\cite{pdg} lists six {\em Cabibbo-favored}  channels,
in addition to two  {\em Cabibbo-suppressed} ones,
whereas
no absolute branching fractions for the $\Xi_c^{0,+}$ decays
have been seen~\cite{pdg}.
Being demonstrated to well fit
the measured values of ${\cal B}(\Lambda_c^+\to {\bf B}_n M)$~\cite{Geng:2017esc},
 $SU(3)_f$ symmetry can be used
to study the $\Xi_c^{0,+}\to {\bf B}_n M$ decays.
For example, according to the data in the PDG~\cite{pdg},
it is given that
\begin{eqnarray}
\frac{{\cal B}(\Xi_c^+\to\Xi^0\pi^+)}{{\cal B}(\Xi_c^+\to\Xi^0 e^+\nu_e)}
&=&0.24\pm 0.11\,,\nonumber\\
\frac{{\cal B}(\Xi_c^0\to\Lambda^0\bar K^0)}{{\cal B}(\Xi_c^0\to\Xi^- e^+\nu_e)}
&=&0.07\pm 0.03\,,
%
\end{eqnarray}
which result in
\begin{eqnarray}\label{B_I}
{\cal B}_I(\Xi_c^+\to\Xi^0\pi^+)&=&(7.2\pm 3.5)\times 10^{-3}\,,\nonumber\\
{\cal B}_I(\Xi_c^0\to\Lambda^0\bar K^0)&=&(8.3\pm 3.7)\times 10^{-3}\,,
\end{eqnarray}
by bringing the predictions of Eq.~(\ref{semi_pred}) into the relations.
On the other hand, the $SU(3)$ parameters for ${\bf B}_{c}\to {\bf B}_n M$
have been extracted from the observed ${\cal B}(\Lambda_c^+\to {\bf B}_n M)$ data,
given by~\cite{Geng:2017esc}
\begin{eqnarray}\label{su3_fit}
&&(a_1,a_2,a_3)=(0.257\pm 0.006,0.121\pm 0.015,0.092\pm 0.021)\,\text{GeV}^3\,,\nonumber\\
&&(\delta_{a_2},\delta_{a_3})=(79.0\pm 6.8, 35.2\pm 8.8)^\circ\,,
\end{eqnarray}
where $\delta_{{a_2},{a_3}}$ are the relative phases from the complex $a_2$ and $a_3$ parameters,
and $a_1$ is fixed to be real.
Besides, we follow Ref.~\cite{Lu:2016ogy}
to ignore $a_{4,5, ...,7}$ from $H(\overline{15})$,
which are based on $(c_-/c_+)^2=5.5$ from $\bar {\cal H}_{eff}$ in Eq.~(\ref{Heff2b}),
leading to the estimation of ${\cal B}(\Lambda_c\to \Sigma^+ K^0)$
with the $(10-15)\%$ deviation from the data~\cite{Geng:2017esc}.
By using  $SU(3)$ parameters in Eq.~(\ref{su3_fit}),
we obtain
\begin{eqnarray}\label{B_II}
{\cal B}_{II}(\Xi_c^+\to\Xi^0\pi^+)&=&(8.0\pm 4.1)\times 10^{-3}\,,\nonumber\\
{\cal B}_{II}(\Xi_c^0\to\Lambda^0\bar K^0)&=&(8.3\pm 0.9)\times 10^{-3}\,.
\end{eqnarray}
In Eqs.~(\ref{B_I}) and (\ref{B_II}),
${\cal B}_{I,II}$ indeed come from  semileptonic and non-leptonic $SU(3)$ relations,
respectively, even though the data inputs have very different sources.
As a result, the good agreements
for $\Xi_c^+\to\Xi^0\pi^+$ and $\Xi_c^0\to\Lambda^0\bar K^0$
clearly support the approach with the $SU(3)_f$ symmetry.

As seen from Table~\ref{tab_Bc1b} for the ${\bf B}_{c}\to {\bf B}'_n M$ decays,
one has that
\begin{eqnarray}\label{Re_A}
&&
{\cal B}(\Lambda_c^+\to \Delta^{++}K^-)=
\frac{1}{s_c^2}{\cal B}(\Lambda_c^+\to \Delta^{++}\pi^-)\nonumber\\
&&
=\frac{1}{s_c^2R_+}{\cal B}(\Xi_c^+\to\Delta^{++}K^-)\nonumber\\
&&
=\frac{3}{s_c^4R_+}{\cal B}(\Xi_c^+\to\Sigma^{\prime+}K^0)
=\frac{1}{s_c^4R_+}{\cal B}(\Xi_c^+\to \Delta^{++}\pi^-)\,,
\end{eqnarray}
and
\begin{eqnarray}\label{Re_B}
&&
{\cal B}(\Xi_c^0\to \Omega^-K^+)=3{\cal B}(\Xi_c^0\to \Xi^{\prime-}\pi^+)\nonumber\\
&&
=\frac{3}{4s_c^2}{\cal B}(\Xi_c^0\to\Sigma^{\prime-}\pi^+,\Xi^{\prime-}K^+)
\nonumber\\
&&
=\frac{1}{s_c^4}{\cal B}(\Xi_c^0\to \Delta^-\pi^+)
=\frac{1}{3s_c^4}{\cal B}(\Xi_c^0\to \Sigma^{\prime-}K^+)\,,
\end{eqnarray}
with $R_{+(0)} =\tau_{\Xi_c^{+(0)}}/\tau_{\Lambda_c^+}$,
whose amplitudes
are commonly
proportional to $2a_8+a_9$ and $2a_8-a_9$, respectively.
Besides, we obtain
\begin{eqnarray}\label{Re_C}
&&
{\cal B}(\Xi_c^0\to \Sigma^{\prime+}K^-)
=
\frac{1}{s_c^2}{\cal B}(\Xi_c^0\to \Delta^{+}K^-,\Sigma^{\prime-}\pi^+)\nonumber\\
&&
=\frac{1}{s_c^4}{\cal B}(\Xi_c^0\to \Delta^{+}\pi^-)
=\frac{1}{2s_c^4}{\cal B}(\Xi_c^0\to \Delta^{0}\pi^0)\,,\nonumber\\
&&
{\cal B}(\Xi_c^+\to \Sigma^{\prime+}\bar K^0,\Xi^{\prime0}\pi^+)
=\frac{R_0}{s_c^4}{\cal B}(\Lambda_c^+\to \Delta^{+}K^0)\,,
\end{eqnarray}
corresponding to $T\propto 2a_8-a_9-2a_{11}$ and $a_{11}$,  respectively.
%
%
Currently,
apart from ${\cal B}(\Lambda_c^+\to \Delta^{++} K^-)$,
it is measured that
${\cal B}(\Xi_c^0\to\Omega^- K^+)=$
$(0.297\pm 0.024)\times {\cal B}(\Xi_c^0\to\Xi^-\pi^+)$~\cite{pdg},
such that we can estimate ${\cal B}(\Xi_c^0\to\Omega^- K^+)$
with the input of ${\cal B}(\Xi_c^0\to\Xi^-\pi^+)=(1.6\pm 0.1)\times 10^{-2}$.
%
Subsequently,
with the the two branching ratios, given by
\begin{eqnarray}\label{2_R}
{\cal B}(\Lambda_c^+\to \Delta^{++} K^-)&=&
(1.09\pm 0.25)\times 10^{-2}~\text{\cite{pdg}}\,,\nonumber\\
{\cal B}(\Xi_c^0\to\Omega^- K^+)&=&
(4.8\pm 0.5)\times 10^{-3}\,,
\end{eqnarray}
and the relations in Eqs.~(\ref{Re_A}) and (\ref{Re_B}),
we predict that
\begin{eqnarray}\label{Re_A_pred}
&&
{\cal B}(\Lambda_c^+\to \Delta^{++}\pi^-)=(5.5\pm 1.3)\times 10^{-3}\,,\nonumber\\
&&
{\cal B}(\Xi_c^+\to\Delta^{++}K^-)=(1.2\pm 0.3)\times 10^{-3}\,,\nonumber\\
&&
{\cal B}(\Xi_c^+\to\Sigma^{\prime+}K^0,\Delta^{++}\pi^-)
=(2.1\pm 0.5,6.2\pm 1.5)\times 10^{-5}\,,
\end{eqnarray}
and
\begin{eqnarray}\label{Re_B_pred}
&&
{\cal B}(\Xi_c^0\to \Xi^{\prime-}\pi^+)
=(1.6\pm 0.2)\times 10^{-3}\,,\nonumber\\
&&
{\cal B}(\Xi_c^0\to\Sigma^{\prime-}\pi^+(\Xi^{\prime-}K^+))
=(3.2\pm 0.3)\times 10^{-4}\,,\nonumber\\
&&
{\cal B}(\Xi_c^0\to \Delta^-\pi^+,\Sigma^{\prime-}K^+)
=(1.2\pm 0.1,3.7\pm 0.4)\times 10^{-5}\,.
\end{eqnarray}
We remark that, if $H(\overline{15})$ is negligible,
one has
${\cal B}(\Xi_c^0\to\Omega^- K^+)\simeq R_0 {\cal B}(\Lambda_c^+\to \Delta^{++} K^-)$
with $R_0=0.56\pm 0.07$, which agrees with the value of $0.44\pm 0.11$
from Eq.~(\ref{2_R}).\\

\noindent
$\bullet$ The ${\bf B}'_{c}\to {\bf B}_n^{(\prime)}M$ decays\\
From Table~\ref{tab_Bc1_c} 
to Table~\ref{tab_Bc1_f},
we show the ${\bf B}'_{c}\to {\bf B}_n^{(\prime)}M$ decays
with ${\bf B}'_{c}=(\Sigma_c,\Xi'_{c},\Omega_c)$.
Experimentally, we have that~\cite{pdg}
\begin{eqnarray}
\frac{{\cal B}(\Omega_c^0\to \Omega^-\pi^+)}{{\cal B}(\Omega_c^0\to \Omega^-e^+\nu_e)}
=0.41\pm 0.19\pm 0.04\,,
\end{eqnarray}
where $\Omega_c^0\to \Omega^-\pi^+$ and $\Omega_c^0\to \Omega^-e^+\nu_e$
are identified from Tables~\ref{tab_semi} and \ref{tab_Bc1_f}
as Cabibbo-allowed processes, with $\Omega^-$
belonging to the baryon decuplet ${\bf B}'_n$.
On the other hand,
as the only Cabibbo-allowed $\Omega_c^0\to {\bf B}_nM$ mode,
$\Omega_c^0\to \Xi^{0} \bar{K}^{0}$ has not been measured yet,
which calls for the other accessible decay modes.
Although it seems that there is no relation for $\Omega_c^0\to {\bf B}_nM$
in Table~\ref{tab_Bc1_d}, if $H(\overline{15})$ is ignorable,
we have
\begin{eqnarray}
{\cal B}(\Omega_c^0\to \Sigma^+ K^-)&=&2{\cal B}(\Omega_c^0\to \Sigma^0 \bar K^0)\,,\nonumber\\
{\cal B}(\Omega_c^0\to \Xi^- \pi^+)&=&2{\cal B}(\Omega_c^0\to \Xi^0 \pi^0)\,,
\end{eqnarray}
for the Cabibbo-suppressed processes, and
\begin{eqnarray}
{\cal B}(\Omega_c^0\to \Sigma^\pm \pi^\mp)&=&
{\cal B}(\Omega_c^0\to \Sigma^0 \pi^0)\,,\nonumber\\
{\cal B}(\Omega_c^0\to \Xi^0 K^0)&=&{\cal B}(\Omega_c^0\to \Xi^- K^+)\,,\nonumber\\
{\cal B}(\Omega_c^0\to p K^-)&=&{\cal B}(\Omega_c^0\to n \bar K^0)\,,
\end{eqnarray}
for the doubly Cabibbo-suppressed ones,
which can be regarded to recover the isospin symmetry.

For $\Omega_c^0\to {\bf B}_n^{\prime}M$,
as seen in Table~\ref{tab_Bc1_f}, it is found that
\begin{eqnarray}
{\cal B}(\Omega_c^0\to \Delta^+ K^-)&=&
{\cal B}(\Omega_c^0\to \Delta^0 \bar K^0)\,,\nonumber\\
{\cal B}(\Omega_c^0\to \Omega^- K^+)&=&
\frac{1}{s_c^2}{\cal B}(\Omega_c^0\to \Xi^{\prime-} K^+)\,.
\end{eqnarray}
In addition,  ignoring $H(\overline{15})$,
we derive the relations with the recovered isospin symmetry, given by
\begin{eqnarray}
{\cal B}(\Omega_c^0\to \Sigma^{\prime+}K^-)&=&
{\cal B}(\Omega_c^0\to \Sigma^{\prime0} \bar K^0)\,,\nonumber\\
{\cal B}(\Omega_c^0\to \Xi^{\prime-}\pi^+)&=&
{\cal B}(\Omega_c^0\to \Xi^{\prime0} \pi^0)\,,
\end{eqnarray}
and
\begin{eqnarray}
{\cal B}(\Omega_c^0\to \Sigma^{\prime\pm}\pi^\mp)&=&
{\cal B}(\Omega_c^0\to \Sigma^{\prime0} \pi^0)\,,\nonumber\\
{\cal B}(\Omega_c^0\to \Xi^{\prime-}K^+)&=&
{\cal B}(\Omega_c^0\to \Xi^{\prime0} K^0)\,,
\end{eqnarray}
for the Cabibbo- and doubly Cabibbo-suppressed decays, respectively.\\

\noindent
$\bullet$ The ${\bf B}_{cc}\to {\bf B}_n^{(\prime)}M_c$ decays\\
For the possible ${\bf B}_{cc}\to {\bf B}_n M_c$ decays  in Table~\ref{tab_B2c_a},
the Cabibbo-allowd decay modes can be related to the (doubly) Cabibbo-suppressed ones,
given by
\begin{eqnarray}
\Gamma(\Xi_{cc}^{++}\to\Sigma^+D^+)&=&
\frac{1}{s_c^2}\Gamma(\Xi_{cc}^+\to pD^+)=
\frac{1}{s_c^4}\Gamma(\Xi_{cc}^+\to pD_s^+)\,,\nonumber\\
\Gamma(\Xi_{cc}^+\to\Sigma^+D^0)&=&
\frac{1}{s_c^2}\Gamma(\Xi_{cc}^+\to pD^0)=
\frac{1}{s_c^4}\Gamma(\Omega_{cc}^+\to pD^0)\,,\nonumber\\
\Gamma(\Xi_{cc}^+\to\Xi^0D_s^+)&=&
\frac{1}{s_c^4}\Gamma(\Omega_{cc}^+\to nD^+)\,,\nonumber\\
\Gamma(\Omega_{cc}^+\to\Xi^0D^+)&=&
\frac{1}{s_c^4}\Gamma(\Xi_{cc}^+\to nD_s^+)\,,\nonumber\\
\Gamma(\Xi_{cc}^+\to\Sigma^0D^+)&=&
\frac{1}{2s_c^2}\Gamma(\Xi_{cc}^+\to nD^+)\,.
\end{eqnarray}
By keeping $b_{1,2}$ from $H(6)$ and
disregarding $b_{3,4}$ from $H(\overline{15})$,
similar to the demonstrations for ${\bf B}_{c}\to {\bf B}_n^{(\prime)}M$,
we obtain additional relations
such as
\begin{eqnarray}
\Gamma(\Xi_{cc}^+\to\Sigma^0D^+)&=&3\Gamma(\Xi_{cc}^+\to\Lambda^0D^+)\,,\nonumber\\
\Gamma(\Xi_{cc}^+\to nD^+)&=&4\Gamma(\Omega_{cc}^+\to \Xi^0 D_s^+)
=\frac{3}{2s_c^4}\Gamma(\Omega_{cc}^+\to\Lambda^0 D_s^+)\,,\nonumber\\
\Gamma(\Omega_{cc}^+\to p D^0)&=&\Gamma(\Omega_{cc}^+\to n D_s^+)\,.
\end{eqnarray}

It is interesting to note that, in contrast with ${\bf B}_{cc}\to {\bf B}_n M_c$,
the ${\bf B}_{cc}\to {\bf B}_n^{\prime}M_c$ decays are suppressed,
where the amplitudes in Eq.~(\ref{T_Bc2_D})
consist of $b_{5,6}$ from $H(\overline{15})$ only,
resulting in contributions 5.5 times smaller than $H(6)$.
According to Table~\ref{tab_B2c_a}, one gets that
\begin{eqnarray}
&&\Gamma(\Xi_{cc}^{++}\to \Sigma^{\prime+}D^+)=
\Gamma(\Omega_{cc}^{+}\to \Xi^{\prime0}D^+)\nonumber\\
&=&\frac{1}{s_c^2}\Gamma(\Xi_{cc}^{++}\to \Delta^+D^+,\Sigma^{\prime+}D_s^+)
=\frac{1}{s_c^4}\Gamma(\Xi_{cc}^{++}\to \Delta^+D_s^+)
=\frac{1}{s_c^4}\Gamma(\Xi_{cc}^{+}\to \Delta^0D_s^+)\,,\nonumber\\
&&\Gamma(\Xi_{cc}^{+}\to \Sigma^{\prime+}D^0)=
\Gamma(\Xi_{cc}^{+}\to \Xi^{\prime0}D_s^+)\nonumber\\
&=&\frac{1}{s_c^2}\Gamma(\Xi_{cc}^{+}\to \Delta^+D^0)
=\frac{1}{s_c^2}\Gamma(\Omega_{cc}^{+}\to \Sigma^{\prime+}D^0)
=\frac{1}{s_c^4}\Gamma(\Omega_{cc}^{+}\to \Delta^+D^0,\Delta^0D^+)\,,
\end{eqnarray}
and
\begin{eqnarray}
&&\Gamma(\Xi_{cc}^{+}\to \Sigma^{\prime0}D^+)\nonumber\\
&=&\frac{1}{2s_c^2}\Gamma(\Xi_{cc}^{+}\to \Delta^0D^+)
=\frac{1}{2s_c^2}\Gamma(\Omega_{cc}^{+}\to \Xi^{\prime0}D_s^+)
=\frac{1}{s_c^4}\Gamma(\Omega_{cc}^{+}\to \Sigma^{\prime0}D_s^+)\,,\nonumber\\
&&\Gamma(\Omega_{cc}^{+}\to \Sigma^{\prime0}D_s^+)
=\Gamma(\Xi_{cc}^{+}\to \Sigma^{\prime0}D^+)\,.
\end{eqnarray}

\noindent
$\bullet$ The ${\bf B}_{cc}\to {\bf B}_{c}^{(\prime)}M$ decays\\
In the ${\bf B}_{cc}\to {\bf B}_{c}M$ decays, the Cabibbo-allowed amplitudes
are composed of $SU(3)$ parameters $b_{7,8}$ 
from $H(6)$, instead of $b_{9,10}$ from $H(\overline{15})$,
which indicate that the decays are measurable.
In fact, the decay mode of $\Xi_{cc}^{++}\to \Xi_{c}^+\pi^+$
has been suggested
to be worth measuring by the model calculation~\cite{Yu:2017zst}.
Here, we connect these Cabibbo-allowed decays to be
\begin{eqnarray}
\Gamma(\Xi_{cc}^{++}\to \Xi_{c}^+\pi^+)&=&
\Gamma(\Omega_{cc}^{+}\to \Xi_{c}^+\bar K^0)\,,\nonumber\\
\Gamma(\Xi_{cc}^{+}\to \Xi_{c}^0\pi^+)&=&
\Gamma(\Xi_{cc}^{+}\to \Lambda_{c}^+\bar K^0)\,,\nonumber\\
\Gamma(\Xi_{cc}^{+}\to \Xi_{c}^+\pi^0)&=&
3\Gamma(\Xi_{cc}^{+}\to \Xi_{c}^+\eta)\,,
\end{eqnarray}
which are the most accessible decay modes to the experiments.
We note that the accuracy of the
prediction involving $\eta$ is limited by the assumption that $\eta$ is a pure octet.
Next, the Cabibbo-suppressed decays are related as
\begin{eqnarray}
\Gamma(\Xi_{cc}^{++}\to \Xi_{c}^+ K^+)&=&
4\Gamma(\Xi_{cc}^{++}\to \Lambda_{c}^+\pi^+)=
8\Gamma(\Xi_{cc}^{++}\to \Lambda_{c}^+\pi^+)\,,\nonumber\\
\Gamma(\Xi_{cc}^{+}\to \Xi_{c}^0 K^+)&=&
4\Gamma(\Omega_{cc}^{+}\to \Xi_{c}^0\pi^+)=
8\Gamma(\Omega_{cc}^{+}\to \Xi_{c}^+\pi^0)\,,\nonumber\\
\Gamma(\Xi_{cc}^{+}\to \Xi_{c}^+ K^0)&=&
4\Gamma(\Omega_{cc}^{+}\to \Lambda_{c}^+\bar K^0)\,.
\end{eqnarray}
For the doubly Cabibbo-suppressed ones,
only when $a_{9,10}$ from $H(\overline{15})$ are negligible,
we can find that
\begin{eqnarray}
\Gamma(\Xi_{cc}^{++}\to \Lambda_{c}^+K^+)&=&
\Gamma(\Xi_{cc}^{+}\to \Lambda_{c}^+K^0)\,,\nonumber\\
\Gamma(\Omega_{cc}^{+}\to \Xi_{c}^0K^+)&=&
\Gamma(\Omega_{cc}^{+}\to \Xi_{c}^+K^0)\,.
\end{eqnarray}
There are three kinds of relations in the ${\bf B}_{cc}\to {\bf B}'_{c}M$ decays,
 given by
\begin{eqnarray}
&&
\Gamma(\Xi_{cc}^{++}\to \Sigma_c^{++}\bar K^0)\nonumber\\
&=&\frac{2}{s_c^2}\Gamma(\Xi_{cc}^{++}\to \Sigma_c^{++}\pi^0)
=\frac{1}{s_c^4}\Gamma(\Xi_{cc}^{++}\to \Sigma_c^{++} K^0)
=\frac{1}{s_c^4}\Gamma(\Xi_{cc}^{+}\to \Sigma_c^{+} K^0)\,,\nonumber\\
&&\Gamma(\Omega_{cc}^+\to \Omega_c^0\pi^+)
=2\Gamma(\Xi_{cc}^{++}\to \Xi_c^{\prime+}\pi^+)\nonumber\\
&=&\frac{2}{s_c^2}\Gamma(\Xi_{cc}^{++}\to \Sigma_c^{+}\pi^+)
=\frac{2}{s_c^4}\Gamma(\Xi_{cc}^{++}\to \Sigma_c^{+}K^+)
=\frac{1}{s_c^4}\Gamma(\Xi_{cc}^{+}\to \Sigma_c^{+}K^0)\,,\nonumber\\
&&\Gamma(\Omega_{cc}^+\to \Xi_c^{\prime+}\bar K^0)
=2\Gamma(\Xi_{cc}^{++}\to \Xi_c^{\prime+}K^+)\,.
\end{eqnarray}
Note that, 
$\Xi_{cc}^{++}\to \Sigma_c^{++}\bar K^{*0}$
with the strong decays of
$\Sigma_c^{++}\to \Lambda_c^+\pi^+$ and $K^{*0}\to K^-\pi^+$,
corresponds to the observation of
$\Xi_{cc}^{++}\to \Lambda_c^+K^-\pi^+\pi^+$~\cite{Yu:2017zst,Aaij:2017ueg}.
Since the vector meson octet ($V$) is nearly the same as
the pseudo-scalar meson ones ($M$) in Eq.~(\ref{B_8}),
the non-leptonic charmed baryon decays with $V$ and $M$
have similar $SU(3)$ amplitudes. Therefore,
as the counterpart of $\Xi_{cc}^{++}\to \Sigma_c^{++}\bar K^{*0}$
observed by LHCb, $\Xi_{cc}^{++}\to \Sigma_c^{++}\bar K^0$ is promising to be observed.
Moreover, with the amplitudes that contain $2a_{14}+2a_{15}$
from $H(6)$ to give larger contributions,
provided that the two terms have a constructive interference,
it is possible that the decays of
$\Xi_{cc}^+\to (\Sigma_c^+\bar K^0,\Xi_c^{\prime0}\pi^+)$
can be more significant than that of $\Xi_{cc}^{++}\to \Sigma_c^{++}\bar K^0$.
\\

\noindent
$\bullet$ ${\bf B}_{ccc}\to {\bf B}_{cc}M$
and ${\bf B}_{ccc}\to {\bf B}_{c}^{(\prime)}M_c$ decays\\
In Table~\ref{tab_Bc3},
the ${\bf B}_{ccc}$ state is indeed the singlet of $\Omega_{ccc}^{++}$,
and the ${\bf B}_{ccc}\to {\bf B}_{cc}M$ decays have two types,
given by
\begin{eqnarray}
&&
\Gamma(\Omega_{ccc}^{++}\to \Xi_{cc}^{++}\bar K^0)\nonumber\\
&=&
\frac{2}{s_c^2}\Gamma(\Omega_{ccc}^{++}\to \Xi_{cc}^{++}\pi^0)=
\frac{2}{3s_c^2}\Gamma(\Omega_{ccc}^{++}\to \Xi_{cc}^{++}\eta)
=\frac{1}{s_c^4}\Gamma(\Omega_{ccc}^{++}\to \Xi_{cc}^{++}K^0)\,,\nonumber\\
&&
\Gamma(\Omega_{ccc}^{++}\to \Omega_{cc}^{+}\pi^+)\nonumber\\
&=&
\frac{1}{s_c^2}\Gamma(\Omega_{ccc}^{++}\to \Xi_{cc}^{+}\pi^+,\Xi_{cc}^{+}K^+)=
\frac{1}{s_c^4}\Gamma(\Omega_{ccc}^{++}\to \Xi_{cc}^{+}K^+)\,,
\end{eqnarray}
where $T$'s are proportional to $d_1-2d_2$ and $d_1+2d_2$, respectively,
with $d_{1(2)}$ from $H(\overline{15}(6))$.
The $\Omega_{ccc}^{++}\to {\bf B}_{c}^{(\prime)}M_c$ decays can be simply related, given by
\begin{eqnarray}
\Gamma(\Omega_{ccc}^{++}\to \Xi_{c}^{+}D^+)&=&
\frac{1}{s_c^2}\Gamma(\Omega_{ccc}^{++}\to \Xi_{c}^{+}D_s^+,\Lambda_{c}^{+}D^+)
=\frac{1}{s_c^4}\Gamma(\Omega_{ccc}^{++}\to \Lambda_{c}^{+}D_s^+)\,,\nonumber\\
\Gamma(\Omega_{ccc}^{++}\to \Xi_{c}^{\prime+}D^+)&=&
\frac{1}{s_c^2}\Gamma(\Omega_{ccc}^{++}\to \Xi_{c}^{\prime+}D_s^+,\Sigma_{c}^{+}D^+)
=\frac{1}{s_c^4}\Gamma(\Omega_{ccc}^{++}\to \Sigma_{c}^{+}D_s^+)\,.
\end{eqnarray}
Note that the decay modes with ${\bf B}_{c}$ and ${\bf B}_{c}^{\prime}$
are in accordance with $d_{4,3}$ from $H(6)$ and $H(\overline{15})$, respectively,
such that it is possible that
the Cabibbo-allowed $\Omega_{ccc}^{++}\to \Xi_{c}^{+}D^+$ decay
can be more accessible to the experiments.

\section{Conclusions}
We have studied the semileptonic and non-leptonic
charmed baryon decays with  $SU(3)_f$ symmetry.
By separating the Cabibbo-allowed decays from
the (doubly) Cabibbo-suppressed ones, we have
provided the accessible decay modes to the experiments
at BESIII and LHCb.
We have predicted the rarely studied
${\bf B}_{c}\to {\bf B}_n^{(\prime)}\ell^+\nu_\ell$ and
${\bf B}_{c}\to {\bf B}_n^{(\prime)}M$ decays, such as
${\cal B}(\Xi_{c}^{0}\to\Xi^{-}e^+\nu_e,\Xi_{c}^{+}\to\Xi^{0}e^+\nu_e)
=(11.9\pm 1.6,3.0\pm 0.5)\times 10^{-2}$,
${\cal B}(\Xi_c^0\to\Lambda^0\bar K^0,\Xi_c^+\to\Xi^0\pi^+)
=(8.3\pm 0.9,8.0\pm 4.1)\times 10^{-3}$,
and ${\cal B}(\Lambda_c^+\to \Delta^{++}\pi^-,\Xi_c^0\to\Omega^- K^+)
=(5.5\pm 1.3,4.8\pm 0.5)\times 10^{-3}$.
We have found that the ${\bf B}_{c}\to {\bf B}'_n\ell^+\nu_\ell$ decays
are forbidden due to the $SU(3)_f$ symmetry. On the other hand,
the $\Omega_c^0\to \Omega^-\ell^+\nu_\ell$,
$\Omega_{cc}^+\to\Omega_c^0\ell^+\nu_\ell$, and
$\Omega_{ccc}^{++}\to \Omega_{cc}^+\ell^+\nu_\ell$ decays have been
presented as the only existing Cabibbo-allowed cases
in
${\bf B}'_{c}\to {\bf B}'_n\ell^+\nu_\ell$, ${\bf B}_{cc}\to {\bf B}'_c\ell^+\nu_\ell$,
and ${\bf B}_{ccc}\to {\bf B}_{cc}^{(\prime)}\ell^+\nu_\ell$,
respectively, where only $\Omega_c^0$ from ${\bf B}'_{c}$ decays weakly.
Moreover, being compatible to
$\Omega_{cc}^{+}\to \Xi_c^+\bar K^0$,
the doubly charmed $\Xi_{cc}^{++}\to \Xi_c^+\pi^+$ decay is
favored to be measured, which agrees with the model calculation.
As the counterpart of $\Xi_{cc}^{++}\to \Sigma_c^{++}\bar K^{*0}$,
which is observed as the resonant
$\Xi_{cc}^{++}\to (\Sigma_c^{++}\to) \Lambda_c^+\pi^+(K^{*0}\to) K^-\pi^+$
four-body decays, $\Xi_{cc}^{++}\to \Sigma_c^{++}\bar K^0$
is promising to be seen.
Finally, the triply
$\Omega_{ccc}^{++}\to (\Xi_{cc}^{++}\bar K^0,\Omega_{cc}^+\pi^+,\Xi_c^+ D^+)$
decays are the favored Cabibbo-allowed decays.


\section*{ACKNOWLEDGMENTS}
This work was supported in part by National Center for Theoretical Sciences,
MoST (MoST-104-2112-M-007-003-MY3), and
National Science Foundation of China (11675030).

\newpage
\begin{table}
\caption{The $T$ amplitudes ($T$-amps) related to
the semileptonic charmed baryon decays.}\label{tab_semi}
{
\scriptsize
\begin{tabular}[t]{|c|c||c|c||c|c|}
\hline
${\bf B}_{c}\to{\bf B}_n$&$T$-amp&
${\bf B}'_{c}\to{\bf B}_n$&$T$-amp&
${\bf B}'_{c}\to{\bf B}'_n$&$T$-amp \\
\hline
$\Xi_{c}^{0}\to\Xi^{-}$&$ \alpha_{1} $&
$ \Xi_{c}'^{0}  \to  \Xi^{-}$ & $-\sqrt{\frac{1}{2}}\alpha_2$ &
$ \Xi_{c}'^{0}  \to  \Xi'^{-}$ & $\sqrt{\frac{2}{3}}\alpha_3$
\\
$\Xi_{c}^{+}\to\Xi^{0}$&$ \alpha_{1}$&
$ \Xi_{c}'^{+}  \to  \Xi^{0}$ & $\sqrt{\frac{1}{2}}\alpha_2$ &
$ \Xi_{c}'^{+}  \to  \Xi'^{0}$ & $\sqrt{\frac{2}{3}}\alpha_3$
\\
$\Lambda_{c}^{+}\to\Lambda^{0}$&$-\sqrt{\frac{2}{3}}\alpha_1$&
&&
&
\\
&&
$ \Sigma_{c}^{0}  \to  \Sigma^{-}$ & $ - \alpha_{2} $ &
$ \Sigma_{c}^{0}  \to  \Sigma'^{-}$ & $\sqrt{\frac{1}{3}}\alpha_3$
\\
&&
$ \Sigma_{c}^{+}  \to  \Sigma^{0}$ & $ - \alpha_{2} $ &
$ \Sigma_{c}^{+}  \to  \Sigma'^{0}$ & $\sqrt{\frac{1}{3}}\alpha_3$
\\
&&
$\Sigma_{c}^{++}  \to  \Sigma^{+} $&$ \alpha_{2} $&
$\Sigma_{c}^{++}\to\Sigma'^{+}$&$\sqrt{\frac{1}{3}}\alpha_3$
\\
&&
&&
$ \Omega_{c}^{0}  \to  \Omega^{-}$ & $ \alpha_{3} $
\\
\hline
$ \Xi_{c}^{0}  \to  \Sigma^{-}$ & $ - \alpha_{1} s_c $ &
$ \Xi_{c}'^{0}  \to  \Sigma^{-}$ & $-\sqrt{\frac{1}{2}}\alpha_2 s_c$ &
$ \Xi_{c}'^{0}  \to  \Sigma'^{-}$ & $-\sqrt{\frac{2}{3}}\alpha_3 s_c$
\\
$ \Xi_{c}^{+}  \to  \Sigma^{0}$ & $\sqrt{\frac{1}{2}}\alpha_1 s_c$&
$ \Xi_{c}'^{+}  \to  \Sigma^{0}$ & $-\frac{1}{2}\alpha_2 s_c$ &
$ \Xi_{c}'^{+}  \to  \Sigma'^{0}$ & $-\sqrt{\frac{1}{3}}\alpha_3s_c$
\\
$ \Xi_{c}^{+}  \to  \Lambda^{0}$ & $-\sqrt{\frac{1}{6}}\alpha_1 s_c$ &
$ \Xi_{c}'^{+}  \to  \Lambda^{0}$ & $-\sqrt{\frac{3}{4}}\alpha_2 s_c$ &
&
\\
$ \Lambda_{c}^{+}  \to  n$ & $ - \alpha_{1} s_c$&
&&
&
\\
&&
&&
$ \Sigma_{c}^{0}  \to  \Delta^{-}$ & $ - \alpha_{3} s_c$
\\
&&
$ \Sigma_{c}^{+}  \to  n$ & $\sqrt{\frac{1}{2}}\alpha_2 s_c$ &
$ \Sigma_{c}^{+}  \to  \Delta^{0}$ & $-\sqrt{\frac{2}{3}}\alpha_3 s_c$
\\
&&
$ \Sigma_{c}^{++}  \to  p$ & $ \alpha_{2} s_c$ &
$ \Sigma_{c}^{++}  \to  \Delta^{+}$ & $-\sqrt{\frac{1}{3}}\alpha_3s_c$
\\
&&
$ \Omega_{c}^{0}  \to  \Xi^{-}$ & $ - \alpha_{2} s_c$ &
$ \Omega_{c}^{0}  \to  \Xi'^{-}$ & $-\sqrt{\frac{1}{3}}\alpha_3s_c$
\\
\hline
\hline
${\bf B}_{ccc}\to{\bf B}_{cc}$&$T$-amp&
${\bf B}_{cc}\to{\bf B}_{c}$&$T$-amp&
${\bf B}_{cc}\to{\bf B}'_{c}$&$T$-amp \\\hline
&&
$ \Xi_{cc}^{+}  \to  \Xi_{c}^{0}$ & $ - \beta_{1} $&
$ \Xi_{cc}^{+}  \to  \Xi_{c}'^{0}$ & $\sqrt{\frac{1}{2}}\beta_2$
\\
&&
$ \Xi_{cc}^{++}  \to  \Xi_{c}^{+}$ & $ \beta_{1} $&
$ \Xi_{cc}^{++}  \to  \Xi_{c}'^{+}$ & $\sqrt{\frac{1}{2}}\beta_2$
\\
$ \Omega_{ccc}^{++}  \to  \Omega_{cc}^{+}$ & $ \delta_{1} $&
&&
$ \Omega_{cc}^{+}  \to  \Omega_{c}^{0}$ & $ \beta_{2} $
\\\hline
&&
&&
$ \Xi_{cc}^{+}  \to  \Sigma_{c}^{0}$ & $ - \beta_{2} s_c$
\\
&&
$ \Xi_{cc}^{++}  \to  \Lambda_{c}^{+}$ & $ \beta_{1} s_c$&
$ \Xi_{cc}^{++}  \to  \Sigma_{c}^{+}$ & $-\sqrt{\frac{1}{2}}\beta_2 s_c$
\\
$ \Omega_{ccc}^{++}  \to  \Xi_{cc}^{+}$ & $ - \delta_{1} s_c$&
$ \Omega_{cc}^{+}  \to  \Xi_{c}^{0}$ & $ - \beta_{1} s_c$&
$ \Omega_{cc}^{+}  \to  \Xi_{c}'^{0}$ & $-\sqrt{\frac{1}{2}}\beta_2 s_c$
\\
\hline
\end{tabular}
}
\end{table}

\begin{table}
\caption{The ${\bf B}_{c}\to {\bf B}_n M$ decays, where
the notations of CA and (D)CS $T$-amps stand for Cabibbo-allowed and
(doubly) Cabibbo-suppressed $T$ amplitudes,
which are the same as those in the following tables.}\label{tab_Bc1a}
{
\scriptsize
\begin{tabular}[t]{|c|l|}
\hline
$\Xi_c^0$&\;\;\;\;\;\;\;\;CA $T$-amp
\\
\hline
$\Sigma^{+} K^{-} $
& $ 2(a_{2}+\frac{a_{4} + a_{7}}{2})$
\\
&
\\
$\Sigma^{0}\bar{K}^{0}$
&$-\sqrt{2}(a_{2}+a_{3}$ 
\\
&$-\frac{a_{6}-a_{7}}{2})$
\\
$\Xi^{0} \pi^{0} $
& $ -\sqrt{2}(a_{1}-a_{3}$
\\
&$-\frac{a_{4}-a_{5}}{2})$
\\
$\Xi^{0} \eta $
& $\sqrt{\frac{2}{3}}(a_1-2a_2-a_3$ 
\\
&$+\frac{a_4+a_5-2a_7}{2})$
\\
$\Xi^{-} \pi^{+} $
& $ 2(a_{1}+\frac{a_{5} + a_{6}}{2})$
\\
$\Lambda^{0} \bar{K}^{0} $
&
$-\sqrt{\frac{2}{3}}(2a_1-a_2-a_3$ 
\\
&$+\frac{2a_5-a_6-a_7}{2})$
\\
&
\\
&
\\
&
\\
&
\\
\hline\hline
$\Xi_c^+$&\;\;\;\;\;\;\;\;CA $T$-amp
\\\hline
&\\
&\\
$\Sigma^{+} \bar{K}^{0} $
&$ -2(a_{3}-\frac{a_{4} + a_{6}}{2})$
\\
&
\\
&\\
&\\
$\Xi^{0} \pi^{+} $
& $2(a_{3}+\frac{a_{4} + a_{6}}{2})$
\\
&
\\
&
\\
&
\\[0.4mm]
\hline\hline
$\Lambda_c^+$&\;\;\;\;\;\;\;\;CA $T$-amp
\\\hline
$\Sigma^{+} \pi^{0} $ & $\sqrt{2}(a_{1}-a_{2}-a_{3}$ 
\\
&$-\frac{a_{5}-a_{7}}{2})$
\\
$\Sigma^{+} \eta $
&
$-\sqrt{\frac{2}{3}}(a_1+a_2-a_3$ 
\\
&$+\frac{2a_4-a_6-a_7}{2})$
\\
$\Sigma^{0} \pi^{+} $
&$-\sqrt{2}(a_1-a_2-a_3$ 
\\
&$-\frac{a_5-a_7}{2})$
\\
$\Xi^{0} K^{+} $
& $-2(a_{2}-\frac{a_{4} + a_{7}}{2})$
\\
$p \bar{K}^{0} $
& $ -2(a_{1}-\frac{a_{5} + a_{6}}{2})$
\\
$\Lambda^{0} \pi^{+} $
&
$-\sqrt{\frac{2}{3}}(a_1+a_2+a_3$
\\
&$-\frac{a_5-2a_6+a_7}{2})$
\\
\hline
\end{tabular}
\begin{tabular}[t]{|c|l|}
\hline
$\Xi_{c}^{0}$ &\;\;\;\;\;\;\;\;\;CS $T$-amp
\\
\hline
$\Sigma^{+} \pi^{-} $
& $-2(a_{2}+\frac{a_{4} + a_{7}}{2})s_c$
\\
$\Sigma^{-} \pi^{+} $
& $-2(a_{1}+\frac{a_{5} + a_{6}}{2})s_c$
\\
$\Sigma^{0} \pi^{0} $
&$-(a_{2}+a_{3}$
\\
&$-\frac{a_{4}-a_{5}+a_{6}-a_{7}}{2})s_c$
\\
$\Xi^{0} K^{0} $
& $2(a_{1}-a_2-a_{3}$
\\
&$+\frac{a_{5}-a_{7}}{2})s_c$
\\
$\Sigma^{0} \eta $
& $\sqrt{\frac{1}{3}}(a_1+a_2+a_3$
\\
&$+\frac{a_4+a_5-3a_6+a_7}{2})s_c$
\\
$\Xi^{-} K^{+} $
& $ -2(a_{1}+\frac{a_{5} + a_{6}}{2})s_c$
\\
$\Lambda^{0} \pi^{0} $
&
$\sqrt{\frac{1}{3}}(a_1+a_2-2a_3$
\\
&$+\frac{a_4-a_5-a_6-a_7}{2})s_c$
\\
$\Lambda^{0} \eta $
&
$(a_1+a_2$
\\
&$-\frac{a_4-a_5+a_6-a_7}{2})s_c$
\\
$n \bar K^{0} $
& $-2(a_{1}-a_{2}-a_{3}$
\\
&$+\frac{a_{5}-a_{7}}{2})s_c$
\\
\hline\hline
$\Xi_{c}^{+}$ &\;\;\;\;\;\;\;\;\;CS $T$-amp
\\\hline
$\Sigma^{0} \pi^{+} $
&$\sqrt 2(a_{1}-a_{2}$
\\
&$+\frac{a_{4}-a_{5}+a_{6}+a_{7}}{2})s_c$
\\
$\Sigma^{+} \pi^{0} $
&$-\sqrt 2(a_{1}-a_{2}$
\\
&$-\frac{a_{4}+a_{5}+a_{6}-a_{7}}{2})s_c$
\\
$\Sigma^{+} \eta$
& $\sqrt{\frac{2}{3}}(a_1+a_2+a_3$
\\
&$-\frac{a_4+a_5+3a_6+a_7}{2})s_c$
\\
$\Xi^{0} K^{+} $
& $2(a_2+a_{3}+\frac{a_{6} - a_{7}}{2})s_c$
\\
$p \bar K^{0} $
& $2(a_1-a_{3}+\frac{a_{4} - a_{5}}{2})s_c$
\\
$\Lambda^0\pi^+$
& $\sqrt{\frac{2}{3}}(a_1+a_2-2a_3$
\\
&$-\frac{3a_4+a_5+a_6+a_7}{2})s_c$
\\
\hline\hline
$\Lambda_{c}^{+}$ &\;\;\;\;\;\;\;\;\;CS $T$-amp
\\\hline
$\Sigma^{+} K^{0} $
& $-2(a_{1}-a_{3}-\frac{a_{4}-a_{5}}{2})s_c$
\\
$\Sigma^{0} K^{+} $
&$-\sqrt{2}(a_1-a_3-\frac{a_4+a_5}{2})s_c$
\\
$p K^-$
&$2(a_{2}+\frac{a_{4} + a_{7}}{2})s_c$
\\
$p \pi^{0} $
& $ -\sqrt 2(a_{2}+a_3-\frac{a_{6} - a_{7}}{2})s_c$
\\
$p \eta $
&
$-\sqrt{\frac{2}{3}}(2a_1-a_2+a_3$
\\
&$+\frac{2a_4+2a_5+3a_6-a_7}{2})s_c$
\\
$\Lambda^{0} K^{+} $
&
$-\sqrt{\frac{2}{3}}(a_1-2a_2+a_3$
\\
&$-\frac{3a_4-a_5+2a_6+2a_7}{2})s_c$
\\
$n\pi^+$
&$-2(a_{2}+a_3-\frac{a_{4} + a_{7}}{2})s_c$
\\
&
\\
\hline
\end{tabular}
\begin{tabular}[t]{|c|l|}
\hline
$\Xi_c^0$&\;\;\;\;\;\;\;\;DCS $T$-amp
\\
\hline
$p\pi^-$&
$-2(a_{2}+\frac{a_{4} + a_{7}}{2})s_c^2$\\
$\Sigma^{-} K^{+} $
& $ 2(a_{1}+\frac{a_{5} + a_{6}}{2})s_c^2$
\\
$\Sigma^{0}{K}^{0}$
&$ \sqrt 2(a_{1}+\frac{a_{5} - a_{6}}{2})s_c^2$
\\
&
\\
$n \pi^{0} $
&$\sqrt 2(a_{2}-\frac{a_{4} - a_{7}}{2})s_c^2$
\\
&
\\
$n\eta $
& $\sqrt{\frac{2}{3}}(2a_1-a_2-2a_3$
\\
&$+\frac{a_4-2a_5+a_7}{2})$
\\
& 
\\
$\Lambda^{0} {K}^{0} $
&
$-\sqrt{\frac{2}{3}}(a_1-2a_2-2a_3$
\\
&$+\frac{a_5+a_6-2a_7}{2})$
\\
&
\\
&
\\
&
\\
&
\\
\hline\hline
$\Xi_c^+$&\;\;\;\;\;\;\;\;DCS $T$-amp
\\\hline
$\Sigma^{0} {K}^{+} $
&$ \sqrt 2(a_{1}-\frac{a_{5} - a_{6}}{2})s_c^2$
\\
$\Sigma^{+} {K}^{0} $
&$ 2(a_{1}-\frac{a_{5} + a_{6}}{2})s_c^2$
\\
$p \pi^0 $
&$ \sqrt 2(a_{2}+\frac{a_{4} - a_{7}}{2})s_c^2$
\\
&
\\
$p\eta$
& $-\sqrt{\frac{2}{3}}(2a_1-a_2-2a_3$
\\
&$-\frac{a_4-2a_5+a_7}{2})s_c^2$
\\
$n \pi^{+} $
& $2(a_{2}-\frac{a_{4} + a_{7}}{2})s_c^2$
\\
&
\\
$\Lambda^0 K^+$
& $\sqrt{\frac{2}{3}}(a_1-2a_2-2a_3$
\\
&$-\frac{a_5+a_6-2a_7}{2})s_c^2$
\\
\hline\hline
$\Lambda_c^+$&\;\;\;\;\;\;\;\;DCS $T$-amp
\\\hline
&
\\
&
\\
&
\\
&
\\
&
\\
&
\\
&
\\
$p {K}^{0} $
& $ 2(a_{3}-\frac{a_{4} + a_{6}}{2})s_c^2$
\\
$nK^+$&
$- 2(a_{3}+\frac{a_{4} + a_{6}}{2})s_c^2$
\\
&
\\[0.4mm]
\hline
\end{tabular}

}

\end{table}

\begin{table}
\caption{The ${\bf B}_{c}\to {\bf B}'_n M$ decays.}\label{tab_Bc1b}
{
\scriptsize
\begin{tabular}[t]{|c|l|}
\hline
$\Xi_{c}^{0}$ &\;\;\;\;\;\;\;\;\;\;CA $T$-amp
\\
\hline
$\Sigma'^{+} K^{-} $
&$\sqrt{\frac{1}{3}}(2a_{8}-a_9-2a_{10})$
\\
$\Sigma'^{0} \bar K^{0} $
&$\sqrt{\frac{2}{3}}(3a_{8}-\frac{a_9}{2}$
\\
&$-a_{10}+a_{11})$
\\
$\Xi'^{-}\pi^+$
&$-\sqrt{\frac{1}{3}}(2a_{8}-a_9)$
\\
&
\\
$\Xi'^{0}\pi^0$
&$-\sqrt{\frac{2}{3}}(a_{8}+\frac{a_9}{2}+2a_{11})$
\\
$\Xi'^{0}\eta$
&$-\sqrt{2}(a_{8}-\frac{a_9}{2}$
\\
&$-\frac{2a_{10}+a_{11}}{3})$
\\
&
\\
$\Omega^- K^+$
&$-(2a_{8}-a_9)$
\\[0.5mm]
\hline\hline
$\Xi_{c}^{+}$ &\;\;\;\;\;\;\;\;\;\;CA $T$-amp
\\\hline
&
\\
&
\\
&
\\
&
\\
$\Sigma'^{+} \bar K^{0} $
&$-\sqrt{\frac{4}{3}}a_{11}$
\\
&
\\
$\Xi'^{0}\pi^+$
&$\sqrt{\frac{4}{3}}a_{11}$
\\[0.5mm]
\hline\hline
$\Lambda_{c}^{+}$ &\;\;\;\;\;\;\;\;\;\;CA $T$-amp
\\\hline
$\Delta^{++} K^{-} $
& $-(2a_{8}+a_9)$
\\
$\Delta^+ \bar K^0$
& $-\sqrt{\frac{1}{3}}(2a_{8}+a_9)$
\\
$\Sigma'^{0}\pi^+$
&$\sqrt{\frac{2}{3}}(a_{8}+\frac{a_9}{2}$\\
&$+a_{10}-a_{11})$
\\
$\Sigma'^{+}  \pi^0$
&$\sqrt{\frac{2}{3}}(a_{8}+\frac{a_9}{2}$
\\
&$+a_{10}-a_{11})$
\\
$\Sigma'^{+}  \eta$
& $\sqrt 2(a_8+\frac{a_9}{2}$
\\
&+$\frac{a_{10}+a_{11}}{3})$
\\
&
\\
$\Xi'^{0} K^+$
&$\sqrt{\frac{1}{3}}(2a_{8}+a_9+2a_{10})$
\\[0.3mm]
\hline
\end{tabular}
\begin{tabular}[t]{|c|l|}
\hline
$\Xi_{c}^{0}$ &\;\;\;\;\;\;\;\;\;\;\;\;\;\;\;\;\;\;\;\;CS $T$-amp
\\
\hline
$\Delta^{+} K^{-} $
& $-\sqrt{\frac{1}{3}}(2a_{8}-a_9-2a_{10})s_c$
\\
$\Delta^{0} \bar K^{0} $
& $-\sqrt{\frac{1}{3}}(2a_{8}-a_9-2a_{10}-2a_{11})s_c$
\\
&
\\
$\Sigma'^{-} \pi^{+} $
&$\sqrt{\frac{4}{3}}(2a_{8}-a_9)s_c$
\\
$\Sigma'^{+} \pi^{-} $
&$-\sqrt{\frac{1}{3}}(2a_{8}-a_9-2a_{10})s_c$
\\
$\Sigma'^{0} \pi^{0} $
&$\;\sqrt{\frac{1}{3}}(3a_{8}-3\frac{a_9}{2}-a_{10}-a_{11})s_c$
\\
$\Sigma'^{0} \eta$
&$(a_{8}-\frac{a_9}{2}-\frac{a_{10}+a_{11}}{3})s_c$
\\
&
\\
$\Xi'^{0}K^0$
&$-\sqrt{\frac{1}{3}}(2a_{8}+a_9+2a_{10}-2a_{11})s_c$
\\
$\Xi'^{-}K^+$
&$\sqrt{\frac{4}{3}}(2a_{8}-a_9)s_c$
\\
\hline\hline
$\Xi_{c}^{+}$ &\;\;\;\;\;\;\;\;\;\;\;\;\;\;\;\;\;\;\;\;CS $T$-amp
\\\hline
$\Delta^{++} K^{-} $
& $(2a_{8}+a_9)s_c$
\\
$\Delta^+\bar K^0$
& $\sqrt{\frac{1}{3}}(2a_{8}+a_9+2a_{11})s_c$
\\
&
\\
$\Sigma'^{0} \pi^{+} $
&$-\sqrt{\frac{2}{3}}(a_{8}+\frac{a_9}{2}+a_{10}+a_{11})s_c$
\\
$\Sigma'^{+} \pi^{0} $
&$-\sqrt{\frac{2}{3}}(a_{8}+\frac{a_9}{2}+a_{10})s_c$
\\
$\Sigma'^{+} \eta$
&$-\sqrt 2(a_{8}+\frac{a_9}{2}+\frac{a_{10}-2a_{11}}{3})s_c$
\\
$\Xi'^{0}K^+$
&$-\sqrt{\frac{1}{3}}(2a_{8}+a_9+2a_{10}-2a_{11})s_c$
\\
\hline\hline
$\Lambda_{c}^{+}$ &\;\;\;\;\;\;\;\;\;\;\;\;\;\;\;\;\;\;\;\;CS $T$-amp
\\\hline
$\Delta^{++} \pi^{-} $
& $(2a_{8}+a_9)s_c$
\\
&
\\
$\Delta^0 \pi^+$
& $-\sqrt{\frac{1}{3}}(2a_{8}+a_9+2a_{10}-2a_{11})s_c$
\\
&
\\
$\Delta^+ \pi^0$
& $-\sqrt{\frac{2}{3}}(2a_{8}+a_9+a_{10}-a_{11})s_c$
\\
&
\\
$\Delta^+ \eta$
& $-{\sqrt 2}(\frac{a_{10}+a_{11}}{3})s_c$
\\
&
\\
$\Sigma'^{+} K^0$
&$\sqrt{\frac{1}{3}}(2a_{8}+a_9+2a_{11})s_c$
\\
$\Sigma'^{0} K^+$
&$-\sqrt{\frac{2}{3}}(2a_{8}+\frac{a_9}{2}+a_{10}+a_{11})s_c$
\\\hline
\end{tabular}
\begin{tabular}[t]{|c|l|}
\hline
$\Xi_{c}^{0}$ &\;\;\;\;\;\;\;\;\;\;DCS $T$-amp
\\
\hline
$\Sigma'^{-} K^{+} $
&$-\sqrt{\frac{1}{3}}(2a_{8}-a_9)s_c^2$
\\
$\Sigma'^{0} K^{0} $
&$\sqrt{\frac{2}{3}}(a_{8}+\frac{a_9}{2}$
\\
&$+\frac{a_{10}-a_{11}}{3})s_c^2$
\\
$\Delta^{-}\pi^+$
&$-(2a_{8}-a_9)s_c^2$
\\
$\Delta^{+}\pi^-$
&$\sqrt{\frac{1}{3}}(2a_{8}-a_9-2a_{10})s_c^2$
\\
$\Delta^{0}\pi^0$
&$-\sqrt{\frac{2}{3}}(2a_{8}-a_9-a_{10})s_c^2$
\\
$\Delta^{0}\eta$
&$-\sqrt{2}(\frac{a_{10}-2a_{11}}{3})s_c^2$
\\
&
\\
&
\\
&
\\[0.5mm]
\hline\hline
$\Xi_{c}^{+}$ &\;\;\;\;\;\;\;\;\;\;DCS $T$-amp
\\\hline
$\Sigma'^{+} K^{0} $
&$-\sqrt{\frac{1}{3}}(2a_{8}+a_9)s_c^2$
\\
$\Sigma'^{0} K^{+} $
&$\sqrt{\frac{1}{6}}(2a_8+a_9$
\\
&$+2a_{10}-2a_{11})s_c^2$
\\
$\Delta^{++}\pi^-$
&$-(2a_{8}+a_9)s_c^2$
\\
$\Delta^{+}\pi^0$
&$\sqrt{\frac{2}{3}}(2a_{8}+a_9+a_{10})s_c^2$
\\
$\Delta^{+}\eta$
&$\sqrt 2(\frac{a_{10}-2a_{11}}{3})s_c^2$
\\
&
\\[0.3mm]
\hline\hline
$\Lambda_{c}^{+}$ &\;\;\;\;\;\;\;\;\;\;DCS $T$-amp
\\\hline
$\Delta^{+} K^{0} $
&$-\sqrt{\frac{4}{3}}a_{11}s_c^2$
\\
$\Delta^0 K^+$
& $\sqrt{\frac{4}{3}}a_{11}s_c^2$
\\
&
\\
&
\\
&
\\
&
\\
&
\\
&
\\
&
\\
&
\\[0.3mm]
\hline
\end{tabular}
}
\end{table}

\begin{table}
\caption{The ${\bf B}'_{c}\to {\bf B}_n M$ decays,
where ${\bf B}'_{c}=(\Sigma^{++},\Sigma^{+},\Sigma^{0})$.}\label{tab_Bc1_c}
{
\scriptsize
		\begin{tabular}[t]{|c|l|}
			\hline
			$ \Sigma_{c}^{++} $&\multicolumn{1}{c|}{CA $T$-amp}\\
			\hline
			$ \Sigma^{+} \pi^{+} $ & $ 2a_{13} + a_{16} + a_{17} $\\
			&\\
			\hline
			\hline
			$ \Sigma_{c}^{+} $&\multicolumn{1}{c|}{CA $T$-amp}\\
			\hline
			$ \Sigma^{+} \pi^{0} $ & $ -a_{13} - a_{14} + a_{15}$\\&$ - a_{16} -\frac{ a_{17}}{2}+\frac{ a_{18}}{2}$\\&$-\frac{ a_{19}}{2}-\frac{ a_{20}}{2} $\\
			$ \Sigma^{+} \eta $ & $ \frac{\sqrt{3}}{6}(2a_{13}+2a_{14}+2a_{15}$\\ &$-3a_{17}+a_{18}-a_{19}-3a_{20})$\\
			$ \Sigma^{0} \pi^{+} $ & $ -a_{13} + a_{14} - a_{15} $\\&$-\frac{ a_{17}}{2}-\frac{ a_{18}}{2}+\frac{ a_{19}}{2}$\\&$+\frac{ a_{20}}{2} $\\
			$ \Xi^{0} K^{+} $ & $ \frac{\sqrt{2}}{2}(2a_{15} +a_{16}+a_{18}+a_{19} )$\\
			$ p \bar{K}^{0} $ & $ \frac{\sqrt{2}}{2}(2a_{14} -a_{20}) $\\
			$ \Lambda^{0} \pi^{+} $ & $ \frac{\sqrt{3}}{6}(2a_{13}+2a_{14}+2a_{15}$\\&$+a_{17}+a_{18}+3a_{19}+a_{20})$\\
			\hline
			\hline
			$ \Sigma_{c}^{0} $&\multicolumn{1}{c|}{CA $T$-amp}\\
			\hline
			$ \Sigma^{+} \pi^{-} $ & $ 2a_{12} + 2a_{15} - a_{16}$\\&$ - a_{19} $\\
			$ \Sigma^{0} \pi^{0} $ & $ 2a_{12} + a_{13} + a_{14} $\\&$+ a_{15} +\frac{ a_{17}}{2}-\frac{ a_{18}}{2}$\\&$-\frac{ a_{19}}{2}+\frac{ a_{20}}{2} $\\
			$ \Sigma^{0} \eta $ & $ -\frac{\sqrt{3}}{6}(2a_{13}+2a_{14}+2a_{15}$\\&$-3a_{17}+a_{18}-a_{19}-3a_{20})$\\
			$ \Sigma^{-} \pi^{+} $ & $ 2a_{12} + 2a_{14} - a_{18} + a_{20} $\\
			$ \Xi^{0} K^{0} $ & $ 2a_{12} + 2a_{15} + a_{16} + a_{19} $\\
			$ \Xi^{-} K^{+} $ & $ 2a_{12} - a_{18} $\\
			$ p K^{-} $ & $ 2a_{12} + a_{18} $\\
			$ n \bar{K}^{0} $ & $ 2a_{12} + 2a_{14} + a_{18} - a_{20} $\\
			$ \Lambda^{0} \pi^{0} $ & $ -\frac{\sqrt{3}}{6}(2a_{13}+2a_{14}+2a_{15}$\\&$+a_{17}+a_{18}+3a_{19}+a_{20})$\\
			$ \Lambda^{0} \eta $ & $ 2a_{12} +\frac{ a_{13}}{3}+\frac{ a_{14}}{3}+\frac{ a_{15}}{3}$\\&$-\frac{ a_{17}}{2}+\frac{ a_{18}}{2}+\frac{ a_{19}}{2}-\frac{ a_{20}}{2} $\\
			\hline
		\end{tabular}
		\begin{tabular}[t]{|c|l|}
			\hline
			$ \Sigma_{c}^{++} $&\multicolumn{1}{c|}{CS $T$-amp}\\
			\hline
			$ \Sigma^{+} K^{+} $ & $( 2a_{13} + a_{16} + a_{17} )s_{c}$\\
			$ p \pi^{+} $ & $( 2a_{13} + a_{16} + a_{17} )s_{c}$\\
			\hline
			\hline
			$ \Sigma_{c}^{+} $&\multicolumn{1}{c|}{CS $T$-amp}\\
			\hline
			$ \Sigma^{+} K^{0} $ & $\frac{ \sqrt{2}}{2}(2a_{13} + 2a_{14} +a_{16}$\\&$-a_{17}-a_{20} )s_{c}$\\
			$ \Sigma^{0} K^{+} $ & $( -a_{13} + a_{14} +\frac{ a_{16}}{2}$\\&$-\frac{ a_{17}}{2}+ a_{19} +\frac{ a_{20}}{2} )s_{c}$\\
			$ p \pi^{0} $ & $( -a_{13} + a_{15} - a_{16} -\frac{ a_{17}}{2}$\\&$+\frac{ a_{18}}{2}-\frac{ a_{19}}{2}- a_{20} )s_{c}$\\
			$ p \eta $ & $ \frac{\sqrt{3}}{6}(2a_{13}-4a_{14}+2a_{15}$\\&$-3a_{17}+a_{18}-a_{19} )s_{c}$\\
			$ n \pi^{+} $ & $\frac{ \sqrt{2}}{2}(2a_{13} + 2a_{15} +a_{17}$\\&$+a_{18}+a_{19} )s_{c}$\\
			$ \Lambda^{0} K^{+} $ & $ \frac{\sqrt{3}}{6}(2a_{13}+2a_{14}-4a_{15}$\\&$-3a_{16}+a_{17}-2a_{18}+a_{20} )s_{c}$\\
			\hline
			\hline
			$ \Sigma_{c}^{0} $&\multicolumn{1}{c|}{CS $T$-amp}\\
			\hline
			$ \Sigma^{0} K^{0} $ & $ \frac{\sqrt{2}}{2}(-2a_{13} -2a_{14} +a_{16}$\\&$+a_{17}+ 2a_{19} +a_{20} )s_{c}$\\
			$ \Sigma^{-} K^{+} $ & $( 2a_{14} + a_{20} )s_{c}$\\
			$ p \pi^{-} $ & $( 2a_{15} - a_{16} - a_{18}$\\&$ - a_{19} )s_{c}$\\
			$ n \pi^{0} $ & $ \frac{\sqrt{2}}{2}(-2a_{13} - 2a_{15} -a_{17}$\\&$+a_{18}-a_{19}- 2a_{20} )s_{c}$\\
			$ n \eta $ & $ \frac{\sqrt{6}}{6}(2a_{13}-4a_{14}+2a_{15}$\\&$-3a_{17}-a_{18}+a_{19} )s_{c}$\\
			$ \Lambda^{0} K^{0} $ & $ \frac{\sqrt{6}}{6}(2a_{13}+2a_{14}-4a_{15}$\\&$-3a_{16}-a_{17}+2a_{18}-a_{20} )s_{c}$\\
			&\\&\\&\\&\\&\\
			\hline
		\end{tabular}
		\begin{tabular}[t]{|c|l|}
			\hline
			$ \Sigma_{c}^{++} $&\multicolumn{1}{c|}{DCS $T$-amp}\\
			\hline
			$ p K^{+} $ & $( 2a_{13} + a_{16} + a_{17} )s_{c}^{2}$\\
			&\\
			\hline
			\hline
			$ \Sigma_{c}^{+} $&\multicolumn{1}{c|}{DCS $T$-amp}\\
			\hline
			$ p K^{0} $ & $ \frac{\sqrt{2}}{2}(2a_{13} +a_{16}-a_{17} )s_{c}^{2}$\\
			$ n K^{+} $ & $ \frac{\sqrt{2}}{2}(2a_{13} -a_{16}+a_{17} ) s_{c}^{2}$\\
			&\\&\\&\\&\\&\\&\\&\\&\\&\\&\\

			\hline
			\hline
			$ \Sigma_{c}^{0} $&\multicolumn{1}{c|}{DCS $T$-amp}\\
			\hline
			$ n K^{0} $ & $( 2a_{13} - a_{16} - a_{17} )s_{c}^{2}$\\
			&\\&\\&\\&\\&\\&\\&\\&\\&\\&\\&\\&\\&\\&\\&\\
			\hline
		\end{tabular}
}
\end{table}

\begin{table}
\caption{The ${\bf B}'_{c}\to {\bf B}_n M$ decays,
where ${\bf B}'_{c}=(\Xi_{c}^{\prime+},\Xi_{c}^{\prime0},\Omega_{c}^{0})$.}\label{tab_Bc1_d}
{
\scriptsize
		\begin{tabular}[t]{|c|l|}
			\hline
			$ \Xi_{c}'^{+} $&\multicolumn{1}{c|}{CA $T$-amp}\\
			\hline
			$ \Sigma^{+} \bar{K}^{0} $ & $ \frac{\sqrt{2}}{2}(2a_{13} +a_{16}-a_{17})$\\
			$ \Xi^{0} \pi^{+} $ & $ \frac{\sqrt{2}}{2}(2a_{13} -a_{16}+a_{17}) $\\
			&\\&\\&\\&\\&\\&\\&\\&\\&\\&\\
			\hline
			\hline
			$ \Xi_{c}'^{0} $&\multicolumn{1}{c|}{CA $T$-amp}\\
			\hline
			$ \Sigma^{+} K^{-} $ & $ \frac{\sqrt{2}}{2}(2a_{15} -a_{16}-a_{18}-a_{19}) $\\
			$ \Sigma^{0} \bar{K}^{0} $ & $ -a_{13} - a_{15} +\frac{ a_{17}}{2}+\frac{ a_{18}}{2}+\frac{ a_{19}}{2} $\\
			$ \Xi^{0} \pi^{0} $ & $ -a_{13} - a_{14} +\frac{ a_{16}}{2}-\frac{ a_{17}}{2}-\frac{ a_{20}}{2} $\\
			$ \Xi^{0} \eta $ & $ \frac{\sqrt{3}}{6}(2a_{13}+2a_{14}-4a_{15}-3a_{16}$\\&$-3a_{17}+2a_{18}-2a_{19}-3a_{20} $\\
			$ \Xi^{-} \pi^{+} $ & $ \frac{\sqrt{2}}{2}(2a_{14} +a_{20}) $\\
			$ \Lambda^{0} \bar{K}^{0} $ & $ \frac{\sqrt{3}}{6}(2a_{13}-4a_{14}+2a_{15}$\\&$-a_{17}-a_{18}+3a_{19}+2a_{20} $\\
			&\\&\\&\\&\\&\\&\\&\\&\\&\\&\\&\\
			\hline
			\hline
			$ \Omega_{c}^{0} $&\multicolumn{1}{c|}{CA $T$-amp}\\
			\hline
			$ \Xi^{0} \bar{K}^{0} $ & $ 2a_{13} - a_{16} - a_{17} $\\
			&\\&\\&\\&\\&\\&\\&\\&\\&\\
			\hline
		\end{tabular}
		\begin{tabular}[t]{|c|l|}
			\hline
			$ \Xi_{c}'^{+} $&\multicolumn{1}{c|}{CS $T$-amp}\\
			\hline
			$ \Sigma^{+} \pi^{0} $ & $( -a_{14} + a_{15} -\frac{ a_{16}}{2}$\\&$- a_{17} +\frac{ a_{18}}{2}-\frac{ a_{19}}{2}-\frac{ a_{20}}{2} )s_{c}$\\
			$ \Sigma^{+} \eta $ & $ \frac{\sqrt{3}}{6}(-4a_{13}+2a_{14}+2a_{15}$\\&$-3a_{16}+a_{18}-a_{19}-3a_{20} )s_{c}$\\
			$ \Sigma^{0} \pi^{+} $ & $( a_{14} - a_{15} -\frac{ a_{16}}{2}$\\&$-\frac{ a_{18}}{2}+\frac{ a_{19}}{2}+\frac{ a_{20}}{2} )s_{c}$\\
			$ \Xi^{0} K^{+} $ & $\frac{\sqrt{2}}{2}(2a_{13} + 2a_{15} +a_{17}$\\&$+a_{18}+a_{19} )s_{c}$\\
			$ p \bar{K}^{0} $ & $\frac{ \sqrt{2}}{2}(2a_{13} + 2a_{14} +a_{16}$\\&$-a_{17}-a_{20} )s_{c}$\\
			$ \Lambda^{0} \pi^{+} $ & $ \frac{\sqrt{3}}{6}(-4a_{13}+2a_{14}+2a_{15}+3a_{16}$\\&$-2a_{17}+a_{18}+3a_{19}+a_{20} )s_{c}$\\
			\hline
			\hline
			$ \Xi_{c}'^{0} $&\multicolumn{1}{c|}{CS $T$-amp}\\
			\hline
			$ \Sigma^{+} \pi^{-} $ & $ \frac{\sqrt{2}}{2}(4a_{12} + 2a_{15} -a_{16}$\\&$+a_{18}-a_{19})s_{c}$\\
			$ \Sigma^{0} \pi^{0} $ & $ \frac{\sqrt{2}}{4}(8a_{12} +2a_{14}+2a_{15}+a_{16}$\\&$+2a_{17}-a_{18}-a_{19}+a_{20} )s_{c}$\\
			$ \Sigma^{0} \eta $ & $\frac{\sqrt{6}}{12}(4a_{13}-2a_{14}-2a_{15}-3a_{16}$\\&$-3a_{18}-3a_{19}+3a_{20} )s_{c}$\\
			$ \Sigma^{-} \pi^{+} $ & $ \frac{\sqrt{2}}{2}(4a_{12} + 2a_{14} - 2a_{18} +a_{20} )s_{c}$\\
			$ \Xi^{0} K^{0} $ & $  \frac{\sqrt{2}}{2}(4a_{12} + 2a_{13} + 2a_{14}+ 2a_{15}$\\&$ -a_{17}+a_{18}+a_{19}-a_{20} )s_{c}$\\
			$ \Xi^{-} K^{+} $ & $ \frac{\sqrt{2}}{2}(4a_{12} + 2a_{14} - 2a_{18} +a_{20} )s_{c}$\\
			$ p K^{-} $ & $\frac{\sqrt{2}}{2}(4a_{12} + 2a_{15} -a_{16}$\\&$+a_{18}-a_{19} )s_{c}$\\
			$ n \bar{K}^{0} $ & $ \frac{\sqrt{2}}{2}(4a_{12} + 2a_{13} + 2a_{14} + 2a_{15}$\\&$ -a_{17}+a_{18}+a_{19}-a_{20} )s_{c}$\\
			$ \Lambda^{0} \pi^{0} $ & $ \frac{\sqrt{6}}{12}(4a_{13}-2a_{14}-2a_{15}$\\&$-3a_{16}-3a_{18}-3a_{19}+3a_{20} )s_{c}$\\
			$ \Lambda^{0} \eta $ & $\frac{ \sqrt{2}}{12}(24a_{12} -8a_{13}+10a_{14}$\\&$+10a_{15}+9a_{16}+6a_{17}$\\&$+3a_{18}+3a_{19}-3a_{20} )s_{c}$\\
			\hline
			\hline
			$ \Omega_{c}^{0} $&\multicolumn{1}{c|}{CS $T$-amp}\\
			\hline
$ \Sigma^{+} K^{-} $
& $( 2a_{15} - a_{16} - a_{18} - a_{19} )s_{c}$\\
$ \Sigma^{0} \bar{K}^{0} $
& $-\sqrt\frac{1}{2}(2a_{15}+a_{16}-a_{18}-a_{19} )s_{c}$\\
$ \Xi^{0} \pi^{0} $
& $-\sqrt\frac{1}{2}(2a_{14}+2a_{17} +a_{20})s_{c}$\\
$ \Xi^{0} \eta $
& $-\sqrt\frac{2}{3} 
(2a_{13}-a_{14}+2a_{15}$\\&$-a_{18}+a_{19}+\frac{ 3}{2} a_{20})s_{c}$\\
			$ \Xi^{-} \pi^{+} $ & $( 2a_{14} + a_{20} )s_{c}$\\
			$ \Lambda^{0} \bar{K}^{0} $ & $ \frac{\sqrt{6}}{3}(-2a_{13}-2a_{14}+a_{15}+\frac{ 3}{2}a_{16}$\\&$+a_{17}-\frac{1}{2}a_{18}+\frac{ 3}{2}a_{19}+a_{20} )s_{c}$\\
			&\\&\\
			\hline
		\end{tabular}
		\begin{tabular}[t]{|c|l|}
			\hline
			$ \Xi_{c}'^{+} $&\multicolumn{1}{c|}{DCS $T$-amp}\\
			\hline
			$ \Sigma^{+} K^{0} $ & $\frac{\sqrt{2}}{2}(2a_{14} -a_{20} )s_{c}^{2}$\\
			$ \Sigma^{0} K^{+} $ & $( a_{14} + a_{19} +\frac{ a_{20}}{2} )s_{c}^{2}$\\
			$ p \pi^{0} $ & $( a_{15} -\frac{ a_{16}}{2}- a_{17} $\\&$+\frac{ a_{18}}{2}-\frac{ a_{19}}{2}- a_{20} )s_{c}^{2}$\\
			$ p \eta $ & $ \frac{\sqrt{3}}{6}(-4a_{13}-4a_{14}+2a_{15}$\\&$- 3a_{16}+a_{18}-a_{19} )s_{c}^{2}$\\
			$ n \pi^{+} $ & $ \frac{\sqrt{2}}{2}(2a_{15} +a_{16}+a_{18}+a_{19} )s_{c}^{2}$\\
			$ \Lambda^{0} K^{+} $ & $\frac{\sqrt{3}}{3}(-2a_{13}+a_{14}-2a_{15}$\\&$-a_{17}-a_{18}+\frac{1}{2}a_{20} )s_{c}^{2}$\\
			&\\&\\&\\
			\hline
			\hline
			$ \Xi_{c}'^{0} $&\multicolumn{1}{c|}{DCS $T$-amp}\\
			\hline
			$ \Sigma^{0} K^{0} $ & $( -a_{14} + a_{19} +\frac{ a_{20}}{2} )s_{c}^{2}$\\
			$ \Sigma^{-} K^{+} $ & $\frac{\sqrt{2}}{2}(2a_{14} +a_{20} )s_{c}^{2}$\\
			$ p \pi^{-} $ & $ \frac{\sqrt{2}}{2}(2a_{15} -a_{16}-a_{18}-a_{19})s_{c}^{2}$\\
			$ n \pi^{0} $ & $( -a_{15} -\frac{ a_{16}}{2}- a_{17}$\\&$ +\frac{ a_{18}}{2}-\frac{ a_{19}}{2}- a_{20} )s_{c}^{2}$\\
			$ n \eta $ & $ \frac{\sqrt{3}}{6}(-4a_{13}-4a_{14}+2a_{15}$\\&$+3a_{16}-a_{18}+a_{19} )s_{c}^{2}$\\
			$ \Lambda^{0} K^{0} $ & $ \frac{\sqrt{3}}{3}(-2a_{13}+a_{14}-2a_{15}$\\&$+a_{17}+a_{18}-\frac{1}{2}a_{20} )s_{c}^{2}$\\
			&\\&\\&\\&\\&\\&\\&\\&\\&\\&\\
			\hline
			\hline
			$ \Omega_{c}^{0} $&\multicolumn{1}{c|}{DCS $T$-amp}\\
			\hline
			$ \Sigma^{+} \pi^{-} $ & $( 2a_{12} + a_{18} )s_{c}^{2}$\\
			$ \Sigma^{0} \pi^{0} $ & $2a_{12}s_{c}^{2}$\\
$ \Sigma^{0} \eta $
& $-\sqrt\frac{1}{3}(a_{18}+2a_{19})s_{c}^{2}$\\
			$ \Sigma^{-} \pi^{+} $ & $( 2a_{12} - a_{18} )s_{c}^{2}$\\
			$ \Xi^{0} K^{0} $ & $( 2a_{12} + 2a_{14} + a_{18} - a_{20} )s_{c}^{2}$\\
			$ \Xi^{-} K^{+} $ & $( 2a_{12} + 2a_{14} - a_{18} + a_{20} )s_{c}^{2}$\\
			$ p K^{-} $ & $( 2a_{12} + 2a_{15} - a_{16} - a_{19} )s_{c}^{2}$\\
			$ n \bar{K}^{0} $ & $( 2a_{12} + 2a_{15} + a_{16} + a_{19} )s_{c}^{2}$\\
$ \Lambda^{0} \pi^{0} $
& $\sqrt\frac{1}{3}(2a_{17}-a_{18}+2a_{20} )s_{c}^{2}$\\
$ \Lambda^{0} \eta $
& $2[ a_{12} +\frac{2}{3}(a_{13}+a_{14}+a_{15})]s_{c}^{2}$\\
\hline
\end{tabular}

}
\end{table}


\begin{table}
\caption{The ${\bf B}'_{c}\to {\bf B}'_n M$ decays,
where ${\bf B}'_{c}=(\Sigma^{++},\Sigma^{+},\Sigma^{0})$.}\label{tab_Bc1_e}
{
\scriptsize
\begin{tabular}[t]{|c|l|}
\hline
$\Sigma_{c}^{++} $&\;\;\;\;\;\;\;\;CA $T$-amp
\\
\hline
$ \Delta^{++} \bar{K}^{0} $ & $ -(2a_{21} - a_{23})$ \\
&\\
$ \Sigma'^{+} \pi^{+} $ & $ \sqrt{\frac{1}{3}}(2a_{21} + a_{23}$\\&$+ 2a_{25}) $
\\[0.6mm]
\hline\hline
$\Sigma_{c}^{+}$&\;\;\;\;\;\;\;\;CA $T$-amp
\\\hline
$ \Delta^{++} K^{-} $ & $-\sqrt{\frac{1}{2}}(2a_{22}-a_{26}) $ \\
$\Delta^+\bar{K^0}$&$-\sqrt{\frac{1}{6}}(4a_{21}+2a_{22}$\\
&$-2a_{23}-a_{26})$ \\
$ \Sigma'^{+} \pi^{0} $ & $\sqrt{\frac{1}{3}}(a_{22}+a_{24}$\\
&$-a_{25}+\frac{a_{26}}{2}) $ \\
$ \Sigma'^{+} \eta $
& $\frac{1}{3}(3a_{22}+a_{24}+a_{25}$\\&$-\frac{a_{26}}{2}) $ \\
$ \Sigma'^{0} \pi^{+} $
& $\sqrt{\frac{1}{3}}(2a_{21}+a_{22}+a_{23}$\\
&$+a_{24}+a_{25}+\frac{a_{26}}{2}) $ \\
$ \Xi'^{0} K^{+} $ & $\sqrt{\frac{2}{3}}(a_{22}+a_{24})$\\&$+6a_{26}$ \\
\hline\hline
$\Sigma_{c}^{0} $&\;\;\;\;\;\;\;\;CA $T$-amp
\\
\hline
$ \Delta^{+} K^{-} $ & $-\sqrt{\frac{1}{3}}(2a_{22}-a_{26}) $ \\
$ \Delta^{0} \bar{K}^{0} $ & $-\sqrt{\frac{1}{3}}(2a_{21}+2a_{22}$\\
&$-a_{23}-a_{26}) $ \\
$ \Sigma'^{+} \pi^{-} $ & $\sqrt{\frac{4}{3}}a_{24}$ \\
$ \Sigma'^{0} \pi^{0} $ & $\sqrt{\frac{4}{3}}(a_{22}-a_{24}$\\
&$-a_{25}+\frac{a_{26}}{2}) $ \\
$ \Sigma'^{0} \eta $ & $ \frac{1}{3}(3a_{22}+a_{24}$\\
&$+a_{25}-\frac{a_{26}}{2}) $ \\
$ \Sigma'^{-} \pi^{+} $ & $\sqrt{\frac{1}{3}}(2a_{21}+2a_{22}$\\
&$+a_{23}+a_{26}) $ \\
$ \Xi'^{0} K^{0} $ & $ \frac{2\sqrt{3}a_{24}}{3} $ \\
$ \Xi'^{-} K^{+} $ & $\sqrt{\frac{1}{3}}(2a_{22}+a_{26}) $ \\
\hline
\end{tabular}
\begin{tabular}[t]{|c|l|}
\hline
$\Sigma_{c}^{++} $&\;\;\;\;\;\;\;\;CS $T$-amp
\\
\hline
$ \Delta^{++} \pi^{0} $ & $-\sqrt{\frac{1}{2}}(2a_{21}-a_{23})s_c$ \\
$ \Delta^{++} \eta $ & $\sqrt{\frac{3}{2}}(2a_{21}-a_{23})s_c$ \\
$ \Delta^{+} \pi^{+} $ & $-\sqrt{\frac{1}{3}}(2a_{21}+a_{23}+2a_{25})s_c$ \\
$ \Sigma'^{+} K^{+} $ & $\sqrt{\frac{1}{3}}(2a_{21}+a_{23}+2a_{25})s_c$ \\
\hline\hline
$\Sigma_{c}^{+}$&\;\;\;\;\;\;\;\;CS $T$-amp\\
\hline
$ \Delta^{++} \pi^{-} $ & $\sqrt{\frac{1}{2}}(2a_{22}-a_{26})s_c$ \\
$ \Delta^{+} \pi^{0} $ & $-\sqrt{\frac{1}{3}}(2a_{21}+2a_{22}-a_{23}$\\
&$+a_{24}-a_{25})s_c$ \\
$ \Delta^{+} \eta $ & $(2a_{21}-a_{23}$\\
&$-\frac{a_{24}+a_{25}+a_{26}}{3})s_c$ \\
$ \Delta^{0} \pi^{+} $ & $\frac{\sqrt{6}}{6}(-4a_{21}-2a_{22}-2a_{23}$\\
&$-2a_{24}-2a_{25}-a_{26})s_c$ \\
$ \Sigma'^{+} K^{0} $ & $\frac{\sqrt{6}}{6}(2a_{22}+2\sqrt{6}a_{25}-a_{26})s_c$ \\
&\\
$ \Sigma'^{0} K^{+} $ & $\frac{\sqrt{3}}{6}(4a_{21}-2a_{22}+2a_{23}$\\
&$-2a_{24}+2a_{25}-a_{26})s_c$ \\
\hline\hline
$\Sigma_{c}^{0} $&\;\;\;\;\;\;\;\;CS $T$-amp\\
\hline
$ \Delta^{+} \pi^{-} $ & $\frac{\sqrt{3}}{3}(2a_{22}-2a_{24}-a_{26})s_c$ \\
$ \Delta^{0} \pi^{0} $ & $\frac{\sqrt{6}}{6}(-2a_{21}-4a_{22}+a_{23}$\\
&$+2a_{24}+2a_{25})s_c$ \\
$ \Delta^{0} \eta $ & $\frac{\sqrt{2}}{6}(6a_{21}-3a_{23}-2a_{24}$\\
&$-2a_{25}-2a_{26})s_c$ \\
$ \Delta^{-} \pi^{+} $ & $( -2a_{21} - 2a_{22} -a_{23} - a_{26} )s_c$ \\
$ \Sigma'^{0} K^{0} $ & $\frac{\sqrt{6}}{6}(2a_{22}-2a_{24}+2a_{25}-a_{26})s_c$ \\
&\\
$ \Sigma'^{-} K^{+} $ & $\frac{\sqrt{3}}{3}(2a_{21}-2a_{22}+a_{23}-a_{26})s_c$ \\
&\\
&\\
&\\
\hline
\end{tabular}
\begin{tabular}[t]{|c|l|}
\hline
$\Sigma_{c}^{++} $&\;\;\;\;\;\;\;\;CDS $T$-amp
\\
\hline
$ \Delta^{++} K^{0} $ & $( 2a_{21}  -a_{23} )s_c^2$ \\
&\\
$ \Delta^{+} K^{+} $ & $-\sqrt{\frac{1}{3}}(2a_{21}+a_{23}$\\&$+2a_{25})s_c^2$
\\[0.6mm]
\hline\hline
$\Sigma_{c}^{+} $&\;\;\;\;\;\;\;\;CDS $T$-amp\\
\hline
&\\
$ \Delta^{+} K^{0} $ & $\sqrt{\frac{2}{3}}(2a_{21}-a_{23}$\\&$-a_{25}) s_c^2$ \\
&\\
&\\
$ \Delta^{0} K^{+} $ & $-\sqrt{\frac{2}{3}}(2a_{21}+a_{23}$\\&$+a_{25})s_c^2$ \\
&\\
&\\
&\\
&\\
\hline\hline
$\Sigma_{c}^{0} $&\;\;\;\;\;\;\;\;CDS $T$-amp\\
\hline
&\\
$ \Delta^{0} K^{0} $ & $\frac{\sqrt{3}}{3}(2a_{21}-a_{23}$\\&$-2a_{25})s_c^2$ \\
&\\
&\\
$ \Delta^{-} K^{+} $ & $-( 2a_{21}+a_{23} )s_c^2$ \\
&\\
&\\
&\\
&\\
&\\
&\\
\hline

\end{tabular}
}
\end{table}

\begin{table}
\caption{The ${\bf B}'_{c}\to {\bf B}'_n M$ decays,
where ${\bf B}'_{c}=(\Xi_{c}^{\prime+},\Xi_{c}^{\prime0},\Omega_{c}^{0})$.}\label{tab_Bc1_f}
{
\scriptsize
\begin{tabular}[t]{|c|l|}
\hline
$\Xi_{c}'^{+} $&\;\;\;\;\;\;\;\;CA $T$-amp\\
\hline
&\\
&\\
$ \Sigma'^{+} \bar{K}^{0} $ & $\frac{\sqrt{6}}{3}(-2a_{21}+a_{23}+a_{25})$ \\
&\\
&\\
&\\
&\\
&\\
$ \Xi'^{0} \pi^{+} $ & $\frac{\sqrt{6}}{3}(2a_{21}+a_{23}+a_{25})$ \\
&\\
\hline\hline
$\Xi_{c}'^{0} $&\;\;\;\;\;\;\;\;CA $T$-amp\\
\hline
&\\
&\\
&\\
$ \Sigma'^{+} K^{-} $ & $ \frac{\sqrt{6}}{6}(-2a_{22}+2a_{24}+a_{26}) $ \\
$ \Sigma'^{0} \bar{K}^{0} $ & $ \frac{\sqrt{3}}{6}(-4a_{21}-2a_{22}+2a_{23}$\\
&$+2a_{24}+2a_{25}+a_{26})$ \\
$ \Xi'^{0} \pi^{0} $ & $ \frac{\sqrt{3}}{6}(2a_{22}-2a_{25}+a_{26}) $ \\
$ \Xi'^{0} \eta $ & $ \frac{1}{6}(6a_{22}-4a_{24}$\\
&$+2a_{25}-a_{26}) $ \\
$ \Xi'^{-} \pi^{+} $ & $ \frac{\sqrt{6}}{6}(4a_{21}+2a_{22}$\\
&$+2a_{23}+a_{26}) $ \\
$ \Omega^{-} K^{+} $ & $ \frac{\sqrt{2}}{2}(2a_{22}+a_{26}) $ \\
\hline\hline
$\Omega_{c}^{0} $&\;\;\;\;\;\;\;\;CA $T$-amp\\
\hline
&\\
&\\
&\\
$ \Xi'^{0} \bar{K}^{0} $ & $-\sqrt{\frac{1}{3}}(2a_{21}-a_{23}-2a_{25}) $ \\
&\\
&\\
&\\
&\\
&\\
$ \Omega^{-} \pi^{+} $ & $ 2a_{21} + a_{23} $ \\
\hline
\end{tabular}
\begin{tabular}[t]{|c|l|}
\hline
$\Xi_{c}'^{+} $&\;\;\;\;\;\;\;\;CS $T$-amp\\
\hline
$ \Delta^{++} K^{-} $ & $\frac{\sqrt{2}}{2}(-2a_{22}+a_{26})s_c$ \\
$ \Delta^{+} \bar{K}^{0} $ & $\frac{\sqrt{3}}{6}(-2a_{22}-2a_{25}+a_{26})s_c$ \\
$ \Sigma'^{+} \pi^{0} $ & $\frac{\sqrt{3}}{6}(-4a_{21}+2a_{22}+2a_{23}$\\
&$+2a_{24}+a_{26})s_c$ \\
$ \Sigma'^{+} \eta $ & $\frac{1}{6}(12a_{21}+6a_{22}-6a_{23}$\\
&$+2a_{24}-4a_{25}-a_{26})s_c$ \\
$ \Sigma'^{0} \pi^{+} $ & $\frac{\sqrt{3}}{6}(-4a_{21}+2a_{22}-2a_{23}$\\
&$+2a_{24}-2a_{25}+a_{26})s_c$ \\
$ \Xi'^{0} K^{+} $ & $\frac{\sqrt{6}}{6}(4a_{21}+2a_{22}+2a_{23}$\\
&$+2a_{24}+2a_{25}+a_{26})s_c$ \\
\hline\hline
$\Xi_{c}'^{0} $&\;\;\;\;\;\;\;\;CS $T$-amp\\
\hline
$ \Delta^{+} K^{-} $ & $\frac{\sqrt{6}}{6}(-2a_{22}-2a_{24}+a_{26})s_c$ \\
$ \Delta^{0} \bar{K}^{0} $ & $\frac{\sqrt{6}}{6}(-2a_{22}-2a_{24}-2a_{25}$\\
&$+a_{26})s_c$ \\
$ \Sigma'^{+} \pi^{-} $ & $\frac{\sqrt{6}}{6}(2a_{22}+2a_{24}-a_{26})s_c$ \\
$ \Sigma'^{0} \pi^{0} $ & $\frac{\sqrt{6}}{6}(-2a_{21}-a_{22}+a_{23}$\\
&$-a_{24}+a_{25}+6a_{26})s_c$ \\
$ \Sigma'^{0} \eta $ & $\frac{\sqrt{2}}{4}(4a_{21}+2a_{22}-2a_{23}$\\
&$+2a_{24}-2a_{25}-a_{26})s_c$ \\
$ \Sigma'^{-} \pi^{+} $ & $\frac{\sqrt{6}}{3}(-2a_{21}-a_{23})s_c$ \\
$ \Xi'^{0} K^{0} $ & $\frac{\sqrt{6}}{6}(2a_{22}+2a_{24}+2a_{25}$\\
&$-a_{26})s_c$ \\
$ \Xi'^{-} K^{+} $ & $\frac{\sqrt{6}}{3}(2a_{21}+a_{23})s_c$ \\
\hline\hline
$\Omega_{c}^{0} $&\;\;\;\;\;\;\;\;CS $T$-amp\\
\hline
$ \Sigma'^{+} K^{-} $
& $-\sqrt{\frac{4}{3}}(a_{22}-a_{24}-\frac{a_{26}}{2})s_c$ \\
$ \Sigma'^{0} \bar{K}^{0} $
& $-\sqrt{\frac{2}{3}}(a_{22}-a_{24}+a_{25}$\\
&$-\frac{a_{26}}{2})s_c$ \\
$ \Xi'^{0} \pi^{0} $
&$-\sqrt{\frac{2}{3}}(a_{21}-a_{22}$\\
&$-\frac{a_{23+a_{26}}}{2})s_c$ \\
$ \Xi'^{0} \eta $
& $\sqrt 2(a_{21}+a_{22}$\\
&$-\frac{3a_{23}+4a_{24}+4a_{25}+a_{26}}{6})s_c$ \\
$ \Xi'^{-} \pi^{+} $
& $-\sqrt{\frac{4}{3}}(a_{21}-a_{22}$\\
&$+\frac{a_{23}-a_{26}}{2})s_c$ \\
$ \Omega^{-} K^{+} $ & $2(a_{21} + a_{22} +\frac{ a_{23} + a_{26}}{2} )s_c$ \\
\hline
\end{tabular}
\begin{tabular}[t]{|c|l|}
\hline
$\Xi_{c}'^{+} $&\;\;\;\;\;\;\;\;CDS $T$-amp\\
\hline
$ \Delta^{++} \pi^{-} $ & $\frac{\sqrt{2}}{2}(2a_{22}-a_{26})s_c^2$ \\
$ \Delta^{+} \pi^{0} $ & $\frac{\sqrt{3}}{3}(-2a_{22}-a_{24})s_c^2$ \\
$ \Delta^{+} \eta $ & $\frac{1}{3}(-a_{24}+2a_{25}-a_{26})s_c^2$ \\
$ \Delta^{0} \pi^{+} $ & $\frac{\sqrt{6}}{6}(-2a_{22}-2a_{24}-a_{26})s_c^2$ \\
$ \Sigma'^{+} K^{0} $ & $\frac{\sqrt{6}}{6}(4a_{21}+2a_{22}-2a_{23}-a_{26})s_c^2$ \\
&\\
$ \Sigma'^{0} K^{+} $ & $\frac{\sqrt{3}}{6}(-4a_{21}-2a_{22}-2a_{23}$\\
&$-2a_{24}-2a_{25}-a_{26})s_c^2$ \\
&\\
&\\

\hline\hline
$\Xi_{c}'^{0} $&\;\;\;\;\;\;\;\;CDS $T$-amp\\
\hline
$ \Delta^{+} \pi^{-} $ & $\frac{\sqrt{6}}{6}(2a_{22}-2a_{24}-a_{26})s_c^2$ \\
$ \Delta^{0} \pi^{0} $ & $\frac{\sqrt{3}}{3}(-2a_{22}+a_{24})s_c^2$ \\
$ \Delta^{0} \eta $ & $\frac{1}{3}(-a_{24}+2a_{25}-a_{26})s_c^2$ \\
$ \Delta^{-} \pi^{+} $ & $\frac{\sqrt{2}}{2}(-2a_{22}-a_{26})s_c^2$ \\
$ \Sigma'^{0} K^{0} $ & $\frac{\sqrt{3}}{6}(4a_{21}+2a_{22}-2a_{23}$\\
&$2a_{24}-2a_{25}-a_{26})s_c^2$ \\
&\\
&\\
$ \Sigma'^{-} K^{+} $ & $\frac{\sqrt{6}}{6}(-4a_{21}-2a_{22}-2a_{23}-a_{26})s_c^2$ \\
&\\
&\\
&\\
\hline\hline
$\Omega_{c}^{0} $&\;\;\;\;\;\;\;\;CDS $T$-amp\\
\hline
$ \Delta^{+} K^{-} $
& $-\sqrt{\frac{4}{3}}a_{24}s_c^2$ \\
$ \Delta^{0} \bar{K}^{0} $
& $-\sqrt{\frac{4}{3}}a_{24}s_c^2$ \\
$ \Sigma'^{+} \pi^{-} $ & $\sqrt{\frac{4}{3}}(a_{22}-\frac{a_{26}}{2})s_c^2$ \\
$ \Sigma'^{0} \pi^{0} $ & $-\sqrt{\frac{4}{3}}a_{22}s_c^2$ \\
$ \Sigma'^{0} \eta $ & $\frac{2}{3}(a_{24}+a_{25}-\frac{a_{26}}{2})s_c^2$ \\
$ \Sigma'^{-} \pi^{+} $ & $-\sqrt{\frac{4}{3}}(a_{22}+\frac{a_{26}}{2})s_c^2$ \\
$ \Xi'^{0} K^{0} $ & $\sqrt{\frac{4}{3}}(a_{21}+a_{22}-\frac{a_{23}+a_{26}}{2})s_c^2$ \\
$ \Xi'^{-} K^{+} $ & $-\sqrt{\frac{4}{3}}(a_{21}+a_{22}+\frac{a_{23}+a_{26}}{2})s_c^2$ \\
&\\
&\\
\hline
\end{tabular}
}
\end{table}

\begin{table}
\caption{The ${\bf B}_{cc}\to {\bf B}_n^{(\prime)} M_c$ decays.}\label{tab_B2c_a}
{
\scriptsize
\begin{tabular}[t]{|c|l|}
\hline
${\bf B}_{cc}\to {\bf B}_nM_c$&\;CA $T$-amp\\
\hline
$ \Xi_{cc}^{++}  \to  \Sigma^{+} D^{+} $
& $ 2b_{2} - b_{4} $ \\
$ \Xi_{cc}^{+}  \to  \Sigma^{+} D^{0} $
& $ 2b_{1} - b_{3} $ \\
$ \Xi_{cc}^{+}  \to  \Sigma^{0} D^{+} $
& $ -\sqrt{2}(b_{1}+b_{2}$\\
&$+\frac{b_{3}+b_4}{2})$ \\
$ \Xi_{cc}^{+}  \to  \Xi^{0} D_{s}^{+} $ & $ 2b_{1} + b_{3} $ \\
$ \Xi_{cc}^{+}  \to  \Lambda^{0} D^{+} $
& $ \sqrt{\frac{2}{3}}(b_{1}+b_2$\\
&$+\frac{b_{3}+b_4}{6})$ \\
$ \Omega_{cc}^{+}  \to  \Xi^{0} D^{+} $ & $ 2b_{2} + b_{4} $ \\
&\\
&\\[0.7mm]
\hline\hline
${\bf B}_{cc}\to {\bf B}_n^{\prime} M_c$&\;CA $T$-amp\\
\hline
$ \Xi_{cc}^{++}  \to  \Sigma'^{+} D^{+} $
& $ 2\sqrt{3}b_{5} $ \\
$ \Xi_{cc}^{+}  \to  \Sigma'^{+} D^{0} $
& $ 2\sqrt{3}b_{6} $ \\
$ \Xi_{cc}^{+}  \to  \Sigma'^{0} D^{+} $
& $ \sqrt{6}(b_{5} +b_{6}) $ \\
$ \Xi_{cc}^{+}  \to  \Xi'^{0} D_{s}^{+} $
& $ 2\sqrt{3}b_{6} $ \\
$ \Omega_{cc}^{+}  \to  \Xi'^{0} D^{+} $
& $ 2\sqrt{3}b_{5} $ \\
&\\
&\\
&\\
\hline
\end{tabular}
\begin{tabular}[t]{|c|l|}
\hline
${\bf B}_{cc}\to {\bf B}_n M_c$&\;\;\;\;\;\;\;\;\;CS $T$-amp\\
\hline
$ \Xi_{cc}^{++}  \to  \Sigma^{+} D_{s}^{+} $
& $-(b_{2} +b_{4} )s_c$\\
$ \Xi_{cc}^{++}  \to  p D^{+} $
& $( 2b_{2} - b_{4} )s_c$ \\
$ \Xi_{cc}^{+}  \to  \Sigma^{0} D_{s}^{+} $
& $\sqrt{\frac{1}{2}}(b_{2}+ 2 b_{3} +b_{4})s_c$\\
$ \Xi_{cc}^{+}  \to  p D^{0} $
& $( 2b_{1} - b_{3} )s_c$\\
$ \Xi_{cc}^{+}  \to  n D^{+} $
& $2(b_{1}+b_{2} +\frac{b_{3}+b_4}{2})s_c$\\
$ \Xi_{cc}^{+}  \to  \Lambda^{0} D_{s}^{+} $
& $-\sqrt{\frac{1}{6}}(4b_1+b_2+3b_4)s_c$\\
$ \Omega_{cc}^{+}  \to  \Sigma^{+} D^{0} $
& $-( b_{1}+b_{3} )s_c$\\
$ \Omega_{cc}^{+}  \to  \Sigma^{0} D^{+} $
& $\sqrt\frac{1}{2}(b_1+b_3+2b_4)s_c$\\
$ \Omega_{cc}^{+}  \to  \Xi^{0} D_{s}^{+} $
& $-(b_{1}+b_{2}-b_{3}-b_{4})s_c$\\
$ \Omega_{cc}^{+}  \to  \Lambda^{0} D^{+} $
&$-\sqrt{\frac{1}{6}}(b_1+4b_2-3b_3)s_c$
\\
\hline\hline
${\bf B}_{cc}\to {\bf B}_n^{\prime} M_c$&\;\;\;\;\;\;\;\;\;CS $T$-amp\\
\hline
$ \Xi_{cc}^{++}  \to  \Delta^{+} D^{+} $
& $-2\sqrt{3}b_{5}s_c$ \\
$ \Xi_{cc}^{++}  \to  \Sigma'^{+} D_{s}^{+} $
& $2\sqrt{3}b_{5}s_c$ \\
$ \Xi_{cc}^{+}  \to  \Delta^{+} D^{0} $
& $-2\sqrt{3}b_{6}s_c$ \\
$ \Xi_{cc}^{+}  \to  \Delta^{0} D^{+} $
& $-2\sqrt{3}( b_{5}+b_{6} )s_c$ \\
$ \Xi_{cc}^{+}  \to  \Sigma'^{0} D_{s}^{+} $
& $ \sqrt{6}(b_{5}-b_{6} )s_c$ \\
$ \Omega_{cc}^{+}  \to  \Sigma'^{+} D^{0} $
& $2\sqrt{3}b_{6}s_c$ \\
$ \Omega_{cc}^{+}  \to  \Sigma'^{0} D^{+} $
& $ -\sqrt{6}(b_{5}-b_{6} )s_c$ \\
$ \Omega_{cc}^{+}  \to  \Xi'^{0} D_{s}^{+} $
& $ 2\sqrt{3}(b_{5}+b_{6} )s_c$ \\
\hline
\end{tabular}
\begin{tabular}[t]{|c|l|}
\hline
${\bf B}_{cc}\to {\bf B}_n M_c$&\;\;\;DCS $T$-amp\\
\hline
$ \Xi_{cc}^{++}  \to  p D_{s}^{+} $
& $( 2b_{2} - b_{4} )s_c^2$\\
$ \Xi_{cc}^{+}  \to  n D_{s}^{+} $
& $( 2b_{2} + b_{4} )s_c^2$\\
$ \Omega_{cc}^{+}  \to  \Sigma^{0} D_{s}^{+} $
& $ \sqrt{2}(b_{3}+b_{4} )s_c^2$\\
$ \Omega_{cc}^{+}  \to  p D^{0} $
& $( 2b_{1} - b_{3} )s_c^2$\\
$ \Omega_{cc}^{+}  \to  n D^{+} $
& $( 2b_{1} + b_{3} )s_c^2$\\
$ \Omega_{cc}^{+}  \to  \Lambda^{0} D_{s}^{+} $
& $-\sqrt{\frac{8}{3}}(b_1+b_2)s_c^2$\\
&\\
&\\
&\\
&\\[0.7mm]
\hline\hline
${\bf B}_{cc}\to {\bf B}_n^{\prime} M_c$&\;\;\;DCS $T$-amp\\
\hline
$ \Xi_{cc}^{++}  \to  \Delta^{+} D_{s}^{+} $
& $-2\sqrt{3}b_{5}s_c^2$ \\
$ \Xi_{cc}^{+}  \to  \Delta^{0} D_{s}^{+} $
& $-2\sqrt{3}b_{5}s_c^2$ \\
$ \Omega_{cc}^{+}  \to  \Delta^{+} D^{0} $
& $-2\sqrt{3}b_{6}s_c^2$ \\
$ \Omega_{cc}^{+}  \to  \Delta^{0} D^{+} $
& $-2\sqrt{3}b_{6}s_c^2$ \\
$ \Omega_{cc}^{+}  \to  \Sigma'^{0} D_{s}^{+} $
& $-\sqrt{6}( b_{5}+b_{6} )s_c^2$ \\
&\\
&\\
&\\[-0.2mm]
\hline
\end{tabular}
}
\end{table}

\newpage
\begin{table}
\caption{The ${\bf B}_{cc}\to {\bf B}_{c}^{(\prime)} M$ decays.}\label{tab_B2c_b}
{
\scriptsize
\begin{tabular}[t]{|c|l|}
\hline
${\bf B}_{cc}\to {\bf B}_{c} M$&\;\;\;\;\;\;CA $T$-amp\\
\hline
$ \Xi_{cc}^{++}  \to  \Xi_{c}^{+} \pi^{+} $
& $ 2b_{7} $ \\
$ \Xi_{cc}^{+}  \to  \Lambda_{c}^{+} \bar{K}^{0} $
& $ 2b_{8} $ \\
$ \Xi_{cc}^{+}  \to  \Xi_{c}^{+} \pi^{0} $
& $ -\sqrt{2}(b_{7}+b_{8})$ \\
$ \Xi_{cc}^{+}  \to  \Xi_{c}^{+} \eta $
& $\sqrt\frac{2}{3}(b_{7}+b_{8})$ \\
$ \Xi_{cc}^{+}  \to  \Xi_{c}^{0} \pi^{+} $
& $ 2b_{8} $ \\
$ \Omega_{cc}^{+}  \to  \Xi_{c}^{+} \bar{K}^{0} $
& $ 2b_{7} $ \\
&\\&\\&\\&\\&\\
&\\[0.7mm]
\hline\hline
${\bf B}_{cc}\to {\bf B}_{c}^{\prime} M$&\;\;\;\;\;\;CA $T$-amp\\
\hline
$ \Xi_{cc}^{++}  \to  \Sigma_{c}^{++} \bar{K}^{0} $
& $ b_{11} + b_{13} - 2b_{14} $ \\
$ \Xi_{cc}^{++}  \to  \Xi_{c}'^{+} \pi^{+} $
& $\sqrt{\frac{1}{2}}(b_{11}+b_{13}$\\
&$+2b_{14})$ \\
$ \Xi_{cc}^{+}  \to  \Sigma_{c}^{++} K^{-} $
& $ b_{12} - 2b_{15} $ \\
$ \Xi_{cc}^{+}  \to  \Sigma_{c}^{+} \bar{K}^{0} $
&$\sqrt{\frac{1}{2}}(b_{11}+b_{12}+b_{13}$\\
&$-2b_{14}-2b_{15})$\\
$ \Xi_{cc}^{+}  \to  \Xi_{c}'^{+} \pi^{0} $
& $ \frac{b_{12}}{2}+ b_{15} $ \\
$ \Xi_{cc}^{+}  \to  \Xi_{c}'^{+} \eta $
& $-\sqrt 3(\frac{b_{12}}{6}-b_{15})$ \\
$ \Xi_{cc}^{+}  \to  \Xi_{c}'^{0} \pi^{+} $
& $\sqrt{\frac{1}{2}}(b_{11}+b_{12}+b_{13}$\\
&$+2b_{14}+2b_{15})$\\
$ \Xi_{cc}^{+}  \to  \Omega_{c}^{0} K^{+} $
& $ b_{12} + 2b_{15} $ \\
$ \Omega_{cc}^{+}  \to  \Xi_{c}'^{+} \bar{K}^{0} $
& $\sqrt{\frac{1}{2}}(b_{11}+b_{13}$\\
&$-b_{14})$\\
$ \Omega_{cc}^{+}  \to  \Omega_{c}^{0} \pi^{+} $
& $ b_{11} + b_{13} + 2b_{14} $ \\
&\\&\\&\\&\\&\\&\\&\\&\\&\\&\\&\\&\\&\\[0.4mm]
\hline
\end{tabular}
\begin{tabular}[t]{|c|l|}
\hline
${\bf B}_{cc}\to {\bf B}_{c} M$&\;\;\;\;\;\;CS $T$-amp\\
\hline
$ \Xi_{cc}^{++}  \to  \Lambda_{c}^{+} \pi^{+} $
& $-b_{7}s_c$\\
$ \Xi_{cc}^{++}  \to  \Xi_{c}^{+} K^{+} $
& $2b_{7}s_c$ \\
$ \Xi_{cc}^{+}  \to  \Lambda_{c}^{+} \pi^{0} $
& $\sqrt{\frac{1}{2}}b_{7}s_c$\\
$ \Xi_{cc}^{+}  \to  \Lambda_{c}^{+} \eta $
& $-\sqrt{\frac{1}{6}}(b_7-4b_8$\\&$+2b_9)s_c$\\
$ \Xi_{cc}^{+}  \to  \Xi_{c}^{+} K^{0} $
& $2(b_{7} +b_{8})s_c$\\
$ \Xi_{cc}^{+}  \to  \Xi_{c}^{0} K^{+} $
& $2b_{8}s_c$\\
$ \Omega_{cc}^{+}  \to  \Lambda_{c}^{+} \bar{K}^{0} $
& $-(b_{7}+b_{8} )s_c$\\
$ \Omega_{cc}^{+}  \to  \Xi_{c}^{+} \pi^{0} $
&$\sqrt{\frac{1}{2}}b_{8}s_c$\\
$ \Omega_{cc}^{+}  \to  \Xi_{c}^{+} \eta $
& $-\sqrt{\frac{2}{3}}(2b_7+\frac{b_{8}}{2}$\\&$ -b_{10})s_c$ \\
$ \Omega_{cc}^{+}  \to  \Xi_{c}^{0} \pi^{+} $
& $-b_{8}s_c$
\\\hline
\hline
${\bf B}_{cc}\to {\bf B}_{c}^{\prime} M$&\;\;\;\;\;\;CS $T$-amp\\
\hline
$ \Xi_{cc}^{++}  \to  \Sigma_{c}^{++} \pi^{0} $
& $\sqrt{\frac{1}{2}}(b_{11}+b_{13}$\\
&$-2b_{14} )s_c$ \\
$ \Xi_{cc}^{++}  \to  \Sigma_{c}^{++} \eta $
& $-\sqrt{\frac{3}{2}}(b_{11}+b_{13})s_c$\\
$ \Xi_{cc}^{++}  \to  \Sigma_{c}^{+} \pi^{+} $
& $-\sqrt{\frac{1}{2}}(b_{11}+b_{13}$\\&$+2b_{14})s_c$\\
$ \Xi_{cc}^{++}  \to  \Xi_{c}'^{+} K^{+} $
& $\sqrt{\frac{1}{2}}(b_{11}+b_{13}$\\&$-b_{14})s_c$\\
$ \Xi_{cc}^{+}  \to  \Sigma_{c}^{++} \pi^{-} $
& $-(b_{12}- 2b_{15} )s_c$\\
$ \Xi_{cc}^{+}  \to  \Sigma_{c}^{+} \pi^{0} $
& $\frac{1}{2}(b_{11}+b_{13}$\\&$-2b_{14} - 4b_{15} )s_c$\\
$ \Xi_{cc}^{+}  \to  \Sigma_{c}^{+} \eta $
& $-\sqrt 3(\frac{b_{11}}{2}+\frac{b_{12}}{3}$\\&$-\frac{b_{13}}{2} )s_c$\\
$ \Xi_{cc}^{+}  \to  \Sigma_{c}^{0} \pi^{+} $
& $-(b_{11}+b_{12}+b_{13}$\\&$+2b_{14}+2b_{15} )s_c$\\
$ \Xi_{cc}^{+}  \to  \Xi_{c}'^{+} K^{0} $
& $-\sqrt{\frac{1}{2}}(b_{12}-2b_{15})s_c$\\
$ \Xi_{cc}^{+}  \to  \Xi_{c}'^{0} K^{+} $
& $\sqrt{\frac{1}{2}}(b_{11}-b_{12}+b_{13}$\\&$-b_{14}-2b_{15})s_c$\\
$ \Omega_{cc}^{+}  \to  \Sigma_{c}^{++} K^{-} $
& $( b_{12} + b_{15} )s_c$\\
$ \Omega_{cc}^{+}  \to  \Sigma_{c}^{+} \bar{K}^{0} $
& $\sqrt{\frac{1}{2}}(b_{12}+b_{15})s_c$\\
$ \Omega_{cc}^{+}  \to  \Xi_{c}'^{+} \pi^{0} $
& $( \frac{b_{11}}{2}+\frac{ b_{12}}{2}+\frac{ b_{13}}{2}$\\&$- b_{14} -\frac{ b_{15}}{2} )s_c$\\
$ \Omega_{cc}^{+}  \to  \Xi_{c}'^{+} \eta $
& $-\sqrt 3(\frac{b_{11}}{2}+\frac{b_{12}}{6}$\\&$+\frac{b_{13}}{2}+\frac{b_{15}}{2})s_c$\\
$\Omega_{cc}^{+}  \to  \Xi_{c}'^{0} \pi^{+} $
& $-\sqrt 2(\frac{b_{11}-b_{12}+b_{13}}{2}$\\&$+b_{14}+\frac{b_{15}}{2} )s_c$\\
$ \Omega_{cc}^{+}  \to  \Omega_{c}^{0} K^{+} $
& $( b_{11} + b_{12} + b_{13}$\\&$ - b_{14} - b_{15} )s_c$\\
\hline
\end{tabular}
\begin{tabular}[t]{|c|l|}
\hline
${\bf B}_{cc}\to {\bf B}_{c} M$&\;\;\;\;\;\;DCS $T$-amp\\
\hline
$ \Xi_{cc}^{++}  \to  \Lambda_{c}^{+} K^{+} $
& $( 2b_{7} - b_{9} )s_c^2$\\
$ \Xi_{cc}^{+}  \to  \Lambda_{c}^{+} K^{0} $
& $( 2b_{7} + b_{9} )s_c^2$\\
$ \Omega_{cc}^{+}  \to  \Lambda_{c}^{+} \eta $
& $-\sqrt{\frac{8}{3}}(b_7+b_8)s_c^2$\\
$ \Omega_{cc}^{+}  \to  \Xi_{c}^{+} K^{0} $
& $(2b_8+b_{10})s_c^2$\\
$ \Omega_{cc}^{+}  \to  \Xi_{c}^{0} K^{+} $
& $( 2b_8-b_{10})s_c^2$\\
&\\&\\&\\&\\&\\&\\&\\[0.5mm]
\hline\hline
${\bf B}_{cc}\to {\bf B}_{c}^{\prime} M$&\;\;\;\;\;\;DCS $T$-amp\\
\hline
$ \Xi_{cc}^{++}  \to  \Sigma_{c}^{++} K^{0} $
& $-(b_{11}+ b_{13}-2b_{14} )s_c^2$\\
$ \Xi_{cc}^{++}  \to  \Sigma_{c}^{+} K^{+} $
& $-\sqrt{\frac{1}{2}}(b_{11}+b_{13}$\\&$+2b_{14})s_c^2$\\
$ \Xi_{cc}^{+}  \to  \Sigma_{c}^{+} K^{0} $
& $-\sqrt{\frac{1}{2}}(b_{11}+b_{13}$\\&$-2b_{14})s_c^2$\\
$ \Xi_{cc}^{+}  \to  \Sigma_{c}^{0} K^{+} $
& $-(b_{11}+b_{13}+2b_{14} )s_c^2$\\
$ \Omega_{cc}^{+}  \to  \Sigma_{c}^{++} \pi^{-} $
& $-(b_{12}-2b_{15} )s_c^2$\\
$ \Omega_{cc}^{+}  \to  \Sigma_{c}^{+} \pi^{0} $
& $-2b_{15}s_c^2$\\
$ \Omega_{cc}^{+}  \to  \Sigma_{c}^{+} \eta $
& $-\sqrt\frac{1}{3}b_{12}s_c^2$\\
$ \Omega_{cc}^{+}  \to  \Sigma_{c}^{0} \pi^{+} $
& $-(b_{12}+2b_{15} )s_c^2$\\
$ \Omega_{cc}^{+}  \to  \Xi_{c}'^{+} K^{0} $
&
$-\sqrt{\frac{1}{2}}(b_{11}+b_{12}+b_{13}$\\&$-2b_{14}-2b_{15})s_c^2$\\
$ \Omega_{cc}^{+}  \to  \Xi_{c}'^{0} K^{+} $
&$-\sqrt{\frac{1}{2}}(b_{11}+b_{12}+b_{13}$\\&$+2b_{14}+2b_{15})s_c^2$\\
&\\&\\&\\&\\&\\&\\&\\&\\&\\&\\&\\&\\&\\[0.7mm]
\hline
\end{tabular}
}
\end{table}

\begin{table}
\caption{The ${\bf B}_{ccc}\to {\bf B}_{cc}M$ and
${\bf B}_{ccc}\to {\bf B}_{c}^{(\prime)} M_c$ decays.}\label{tab_Bc3}
{
\scriptsize
\begin{tabular}[t]{|c|l|}
\hline
${\bf B}_{ccc}\to {\bf B}_{cc} M$&CA $T$-amp\\
\hline
$ \Omega_{ccc}^{++}  \to  \Xi_{cc}^{++} \bar{K}^{0} $
& $ d_{1} - 2d_{2} $ \\
$ \Omega_{ccc}^{++}  \to  \Omega_{cc}^{+} \pi^{+} $
& $ d_{1} + 2d_{2} $ \\
&\\
&\\[0.7mm]
\hline\hline
${\bf B}_{ccc}\to {\bf B}_{c}^{(\prime)} M_c$&CA $T$-amp\\
\hline
$ \Omega_{ccc}^{++}  \to  \Xi_{c}^{+} D^{+} $
& $ 2d_{4} $ \\
$ \Omega_{ccc}^{++}  \to  \Xi_{c}'^{+} D^{+} $
& $ \sqrt{2}d_{3} $ \\
&\\&\\
\hline
\end{tabular}
\begin{tabular}[t]{|c|l|}
\hline
${\bf B}_{ccc}\to {\bf B}_{cc} M$&\;\;\;\;CS $T$-amp\\
\hline
$ \Omega_{ccc}^{++}  \to  \Xi_{cc}^{++} \pi^{0} $
& $\sqrt{\frac{1}{2}}(d_{1}-2d_{2} )s_c$\\
$ \Omega_{ccc}^{++}  \to  \Xi_{cc}^{++} \eta $
& $\sqrt{\frac{3}{2}}(d_{1}-2d_{2} )s_c$ \\
$ \Omega_{ccc}^{++}  \to  \Xi_{cc}^{+} \pi^{+} $
& $-(d_{1}+2d_{2} )s_c$\\
$ \Omega_{ccc}^{++}  \to  \Omega_{cc}^{+} K^{+} $
& $( d_{1} + 2d_{2} )s_c$\\[0.3mm]
\hline\hline
${\bf B}_{ccc}\to {\bf B}_{c}^{(\prime)} M_c$&\;\;\;\;CS $T$-amp\\
\hline
$ \Omega_{ccc}^{++}  \to  \Xi_{c}^{+} D_{s}^{+} $
& $2d_{4}s_c$\\
$ \Omega_{ccc}^{++}  \to  \Lambda_{c}^{+} D^{+} $
& $2d_{4}s_c$ \\
$ \Omega_{ccc}^{++}  \to  \Xi_{c}'^{+} D_{s}^{+} $
& $\sqrt{2}d_{3}s_c$\\
$ \Omega_{ccc}^{++}  \to  \Sigma_{c}^{+} D^{+} $
& $-\sqrt{2}d_{3}s_c$\\
\hline
\end{tabular}
\begin{tabular}[t]{|c|l|}
\hline
${\bf B}_{ccc}\to {\bf B}_{cc} M$&\;\;DCS $T$-amp\\
\hline
$ \Omega_{ccc}^{++}  \to  \Xi_{cc}^{++} K^{0} $
& $-(d_{1}-2d_{2})s_c^2$\\
$ \Omega_{ccc}^{++}  \to  \Xi_{cc}^{+} K^{+} $
& $-(d_{1}+2d_{2})s_c^2$\\
&\\
&\\[0.7mm]
\hline\hline
${\bf B}_{ccc}\to {\bf B}_{c}^{(\prime)} M_c$&\;DCS $T$-amp\\
\hline
$ \Omega_{ccc}^{++}  \to  \Lambda_{c}^{+} D_{s}^{+} $
& $ 2d_{4}s_c^2$\\
$ \Omega_{ccc}^{++}  \to  \Sigma_{c}^{+} D_{s}^{+} $
& $ -\sqrt{2}d_{3}s_c^2$\\
&\\
&\\[-0.2mm]
\hline
\end{tabular}
}
\end{table}

\end{document}